\documentstyle[epsfig,12pt,preprint,tighten,aps]{revtex}
\begin{document}

\draft

\title{\rightline{{\tt July 2000}}
\rightline{{\tt UM-P-029/2000}} \rightline{{\tt RCHEP-005/2000}} \ \\
Further studies on relic neutrino asymmetry generation I: \\ the
adiabatic Boltzmann limit, non-adiabatic evolution, \\
 and the  classical harmonic
oscillator analogue of the \\ quantum kinetic equations}
\author{Raymond R. Volkas and Yvonne Y. Y. Wong}
\address{School of Physics\\
Research Centre for High Energy Physics\\ The University of
Melbourne Vic 3010\\ Australia\\ (r.volkas@physics.unimelb.edu.au,
ywong@physics.unimelb.edu.au)}

\maketitle

\begin{abstract}
We demonstrate that the relic neutrino asymmetry evolution
equation derived from the quantum kinetic equations (QKEs) reduces
to the Boltzmann limit that is dependent only on the instantaneous
neutrino distribution functions, in the adiabatic limit in
conjunction with sufficient damping.  An original physical and/or
geometrical interpretation of the adiabatic approximation is
given, which serves as a convenient visual aid to understanding
the sharply contrasting resonance behaviours exhibited by the
neutrino ensemble in opposing collision regimes.  We also present
a classical analogue for the evolution of the difference in the
$\nu_{\alpha}$ and $\nu_s$ distribution functions which, in the
Boltzmann limit, is akin to the behaviour of the generic reaction
$A \rightleftharpoons B$ with equal forward and reverse reaction
rate constants.  A new characteristic quantity, the matter and
collision-affected mixing angle of the neutrino ensemble, is
identified here for the first time. The role of collisions is
revealed to be twofold: (i) to wipe out the inherent oscillations,
and (ii) to equilibrate the $\nu_{\alpha}$ and $\nu_s$
distribution functions in the long run.  Studies on non-adiabatic
evolution and its possible relation to rapid oscillations in
lepton number generation also feature, with the introduction of an
adiabaticity parameter for collision-affected oscillations.
\end{abstract}

\section{Introduction}

The effects of decohering collisions and coherent flavour
oscillations on a neutrino ensemble are collectively quantified by
the quantum kinetic equations (QKEs) \cite{qke,qke2} which, in
recent years, have been frequently applied in the study of
neutrino asymmetry (difference in neutrino and antineutrino number
densities) evolution in the early universe, with appreciable
success  \cite{longpaper,p&r,thelot}.  In essence, the simplest
scenario involves independent oscillations of an active neutrino
$\nu_{\alpha}$ and its antineutrino $\overline{\nu}_{\alpha}$,
commencing with equal number densities, to a corresponding,
initially absent light sterile species, $\nu_s$ and
$\overline{\nu}_s$, in an environment that alters the two sets of
oscillation parameters in dissimilar ways.  Both $\nu_{\alpha}
\leftrightarrow \nu_s$ and $\overline{\nu}_{\alpha}
\leftrightarrow \overline{\nu}_s$ systems evolve subject
simultaneously to a biased collision scheme which, in crude terms,
is blind to the sterile variety.  For the appropriate vacuum
oscillation parameters, numerical solutions to the exact QKEs
demonstrate the combined effect to be one that sees the relic
$\alpha$-neutrino asymmetry grow to orders of magnitude above the
baryon--antibaryon asymmetry \cite{longpaper,p&r,thelot,ftv}. For
other relevant works see Ref.\ \cite{bigbunch}.\footnote{The magnitude
of the final asymmetry found in Refs.\ \cite{longpaper,thelot} was
recently questioned in Ref.\ \cite{dolgov}. Reference \cite{comment}
will explain the nature of the error in Ref.\ \cite{dolgov}.}

As much as one would like to use the QKEs directly in all
applications, obtaining exact numerical solutions remains a
computationally intensive task, given the necessity to track
neutrinos at all momenta where nonzero feedback couples their
development. Several approximate treatments have been employed in
the past which, besides lifting the burden on the computer
considerably, also on occasion offer valuable analytical insights
on the nature of the asymmetry evolution \cite{longpaper,ftv,bvw}.
Two such schemes are the well-established static approximation
\cite{longpaper,ftv}, and the closely related
adiabatic limit approximation introduced in
Ref.\ \cite{bvw}.

Beginning with the QKEs, the adiabatic limit approach comprises a
set of systematic approximations, leading ultimately to an
approximate expression for the neutrino asymmetry evolution. The
extraction procedure, however,  was hitherto largely motivated by
mathematical convenience; physical, or at least geometrical,
interpretations of the intermediate steps and quantities that
arose therein were lacking. It is thus our intention in this paper
to examine the derivation of the adiabatic approximation more
closely, and to assign definite physical meanings to as many
mathematical man{\oe}uvres as possible.  These new interpretations
are most useful for visually tracking the evolution of the
neutrino ensemble across a Mikheyev--Smirnov--Wolfenstein (MSW)
resonance \cite{msw} in different collision regimes, and in
understanding the nature of non-adiabatic evolution.  We shall
also demonstrate again, this time more assertively, that the
adiabatic procedure leads to the elimination of phase dependence
in the regime of interest: the approximate evolution equations for
the neutrino distribution functions depend only on the
distribution functions themselves, and {\it not} on the coherence
history of the ensemble. In other words, the QKEs yield
classical-like Boltzmann equations in the adiabatic limit.

Our second goal in the present work is to draw attention to the
similarity between some aspects of the QKEs and the behaviours of
the more familiar classical linear harmonic oscillator.  In
particular, the evolution of the difference between the
$\nu_{\alpha}$ and $\nu_s$ distribution functions in the adiabatic
limit may be modelled by a damped harmonic oscillator with a
decaying oscillation midpoint. This analogy turns out to be a very
illuminating one that illustrates clearly the reason behind the
remarkable accuracy and the conceptual correctness of the original
heuristic static approximation.  A new quantity, the neutrino
ensemble's matter and collision-affected mixing angle, is also
established in the course.

Before proceeding, let us state plainly what we hope to achieve
ultimately from these abstract analyses: Acquisition of a clear
understanding of the physical processes that constitute the
various computationally convenient approximations will leave us
better equipped to improve on them. To this end, classical
analogies are especially useful as visual aids.  On a grander
scale, studies of the QKEs {\it per se} may stand to benefit areas
beyond relic neutrino asymmetry evolution, most notably
transitions in multi-level atomic systems which are described by
similar equations \cite{atomic}.

The structure of this paper is as follows: The exact QKEs are
presented in Sec.\ \ref{definitions} for the purpose of
introducing the nomenclature.  Section \ref{adiabatic} is devoted
to the discussion of the formal adiabatic procedure, in which we
shall also present results from numerically integrating the
pertinent approximate neutrino asymmetry evolution equation for
the first time.  The classical harmonic oscillator analogy is to
be treated in Sec.\ \ref{classical}, while Sec.\
\ref{nonadiabaticity} deals with the issue of non-adiabaticity and
its possible relation to rapid oscillations in the asymmetry
evolution. We conclude in Sec.\ \ref{conclusion}.

\section{Neutrino asymmetry evolution and
the quantum kinetic equations: nomenclature}
\label{definitions}

We shall consider a two-flavour system consisting of an active species
$\nu_{\alpha}$ (where $\alpha = e$, $\mu$ or $\tau$), and a
sterile species $\nu_s$, where their respective abundances and
the ensemble's  coherence status at momentum $p$ are encoded in
the density matrices \cite{qke,qke2}
\begin{equation}
\rho(p) = \frac{1}{2} [ P_0(p) + {\bf P}(p) \cdot \sigma ].
\end{equation}
The vector ${\bf P}(p) = P_x(p) \hat{x} + P_y(p) \hat{y} + P_z(p)
\hat{z}$ may be interpreted as the ``polarisation'' and $\sigma =
\sigma_x \hat{x} + \sigma_y \hat{y} + \sigma_z \hat{z}$ are the
Pauli matrices.  In this notation, the $\nu_{\alpha}$ and $\nu_s$
distribution functions at $p$ are respectively
\begin{eqnarray}
N_{\alpha}(p) &=& \frac{1}{2} [P_0(p) + P_z(p)]N^{\text{eq}}(p,0),
\nonumber \\ N_s(p) &=&\frac{1}{2} [P_0(p) -
P_z(p)]N^{\text{eq}}(p,0),
\end{eqnarray}
for which we have chosen the reference distribution function
$N^{\text{eq}}(p,0)$ to be of Fermi--Dirac (equilibrium) form,
\begin{equation}
N^{\text{eq}}(p,\mu) = \frac{1}{2 \pi^2}
\frac{p^2}{1+\exp\left(\frac{p-\mu}{T}\right)},
\end{equation}
with chemical potential $\mu$ set to zero at temperature $T$.  The
four variables $P_0(p)$ and ${\bf P}(p)$ advance in time according
to the quantum kinetic equations (QKEs)
\begin{eqnarray}
\label{qkes}
\frac{\partial {\bf P}}{\partial t} &=& {\bf V}(p)
\times {\bf P}(p) - D(p)[P_x(p) \hat{x} + P_y(p) \hat{y}] +
\frac{\partial P_0}{\partial t} \hat{z},\nonumber
\\ \frac{\partial P_0}{\partial t} &\simeq& \Gamma(p) \left\{
\frac{N^{\text{eq}}(p,\mu)}{N^{\text{eq}}(p,0)} -
\frac{1}{2}[P_0(p) + P_z(p)] \right\},
\end{eqnarray}
where the quantities $D(p)=\Gamma(p)/2$ and ${\bf V}(p)$ {\it
individually} characterise the collision-induced decohering and
the matter-affected coherent aspects of the ensemble's evolution
respectively.\footnote{It shall be demonstrated later that the
system's damping and oscillatory features are in fact mutually
dependent.}  Their approximate forms shall be detailed shortly.
Note that the $\frac{\partial P_0}{\partial t}$ equation is not
exact because the right hand side assumes thermal equilibrium for
all species in the background plasma, while the $\nu_{\alpha}$
distribution is taken to be approximately thermal \cite{bvw}. The properties
of the antineutrino ensemble may be parameterised in a similar
manner and subject to the same QKEs with the appropriate
decoherence function and matter potential. Henceforth, all
quantities pertaining to the $\overline{\nu}_{\alpha}
\leftrightarrow \overline{\nu}_s$ system shall bear an overhead
bar.

The vector ${\bf V}(p)$ reads \cite{rn}
\begin{equation}
\label{potential} {\bf V}(p) = \beta(p) \hat{x} + \lambda(p)
\hat{z},
\end{equation}
with
\begin{eqnarray}
\label{betalambda}
\beta (p) &=& \frac{\Delta m^2}{2p} \sin 2
\theta_0, \nonumber\\ \lambda (p) &=& \frac{\Delta m^2}{2p}[b(p) -
a(p) - \cos 2 \theta_0],
\end{eqnarray}
in which $\Delta m^2$ is the mass-squared difference between the
neutrino states, $\theta_0$ is the vacuum mixing angle, and
\begin{eqnarray}
\label{ap} a(p) &=& -\frac{4 \zeta(3) \sqrt{2} G_F L^{(\alpha)}
T^3 p}{\pi^2 \Delta m^2}, \nonumber \\ b(p) &=& -\frac{4 \zeta(3)
\sqrt{2} G_F A_{\alpha} T^4 p^2}{\pi^2 \Delta m^2 m^2_W},
\end{eqnarray}
given that $G_F$ is the Fermi constant, $m_W$ the $W$-boson mass,
$\zeta$ the Riemann zeta function and $A_e \simeq 17$,
$A_{\mu,\,\tau} \simeq 4.9$.  The quantity
\begin{equation}
\label{efflep}
L^{(\alpha)} = L_{\alpha}+L_e + L_{\mu} + L_{\tau}
+ \eta \equiv 2L_{\alpha} + \tilde{L},
\end{equation}
combines the various asymmetries individually defined as
\begin{equation}
L_{\alpha} = \frac{n_{\alpha} -
\overline{n}_{\alpha}}{n_{\gamma}},
\end{equation}
where $n_{\alpha}= \int N_{\alpha}(p) dp$, $n_\gamma$ is the
photon number density, and $\eta$ is a small term related to the
cosmological baryon--antibaryon asymmetry.  The authors of Ref.\
\cite{p&r} have called $L^{(\alpha)}$ the {\it effective total
lepton number} (for the $\alpha$-neutrino species), a name we
shall also adopt.\footnote{The terms ``lepton number'' and
``neutrino asymmetry'' are used interchangeably throughout this
work.} The condition $\lambda = 0$ is identified with a
Mikheyev--Smirnov--Wolfenstein (MSW) resonance \cite{msw}.

The function $D(p)$ is equivalent to half of the total collision
rate for $\nu_{\alpha}$ with momentum $p$, that is \cite{qke,bvw},
\begin{equation}
\label{dp}
 D(p)=\frac{1}{2} \Gamma(p) = \frac{1}{2}
\frac{p}{\langle p \rangle_0} [y_{\alpha} G^2_F T^5 (1 -
z_{\alpha} L_{\alpha}) + {\cal O}(L^2_{\alpha})],
\end{equation}
where $\langle p \rangle_0 \simeq 3.15 T$ is the average momentum
for a relativistic Fermi--Dirac distribution with zero chemical
potential, $y_e \simeq 4$, $y_{\mu, \, \tau} \simeq 2.9$ and $z_e
\simeq 0.1$, $z_{\mu,\,\tau} \simeq 0.04$.

The corresponding functions $\overline{\bf V}(p)$ and
$\overline{D}(p)$ for the antineutrino system are obtained from
their ordinary counterparts by replacing $L^{(\alpha)}$ and
$L_{\alpha}$ with $-L^{(\alpha)}$  and $-L_{\alpha}$ in Eqs.\
(\ref{ap}) and (\ref{dp}) respectively.

We conclude this section by noting that a direct time evolution
equation for the neutrino asymmetry may be derived from the QKEs
together with $\alpha + s$ lepton number conservation \cite{bvw},
which reads
\begin{equation}
\label{exactdldt}
\frac{dL_{\alpha}}{dt} = \frac{1}{2n_{\gamma}}
\int \beta [P_y(p) - \overline{P}_y(p)] N^{\text{eq}}(p,0) dp.
\end{equation}
As well as serving as the backbone on which to develop useful
approximations,  this expression is useful when numerically
solving the QKEs. Although it is redundant, Eq.\ (\ref{exactdldt})
has the virtue of tracking the crucial quantity $L_{\alpha}$
without the need to take the difference of two large numbers.

\section{Adiabatic limit}
\label{adiabatic}

\subsection{The Boltzmann limit}
\label{boltzmannlimit}

The adiabatic limit approximation introduced in of Ref.\
\cite{bvw}\footnote{Note that the derivation we now review
is very similar to a procedure described in Ref.\ \cite{burgess},
though the context is different.}
 consists of firstly setting the repopulation function
in Eq.\ (\ref{qkes}) to zero, i.e., $\frac{\partial P_0}{\partial
t} \simeq 0$, such that the QKEs simplify to the homogeneous
equations
\begin{equation}
\label{pqke}
\frac{\partial}{\partial t} \left( \begin{array}{c}
                                    P_x \\
                                    P_y \\
                                    P_z \end{array} \right)
\simeq \left( \begin{array}{ccc}
            -D & -\lambda & 0 \\
            \lambda & -D & -\beta \\
            0 & \beta & 0 \end{array} \right)
            \left( \begin{array}{c}
                    P_x \\
                    P_y \\
                    P_z \end{array} \right)
\equiv {\cal K} {\bf P},
\end{equation}
where the dependent variables and coefficients are understood to
be functions of both time and momentum.\footnote{The approximation
$\frac{\partial P_0}{\partial t} \simeq 0$ requires careful
justification, although past numerical evidence has always
strongly suggested this idealisation to be very reasonable.
One can derive in the adiabatic limit of the full-fledged QKEs, including
the finite repopulation term, results
identical to those obtained in this section to leading order.
See the companion paper Ref.\ \cite{lvw}.}

We solve Eq.\ (\ref{pqke})
by establishing a parameter-dependent ``instantaneous diagonal
basis'' ${\cal S}_q \equiv {\cal S}_q(D,\,\lambda,\, \beta)=(
\begin{array}{ccc} \hat{q}_1, & \hat{q}_2, & \hat{q}_3 \end{array})$,
onto which we map the vector ${\bf P}$ from its original ``fixed''
coordinate system ${\cal S} = (\begin{array}{ccc} \hat{x}, &
\hat{y}, & \hat{z}
\end{array})$ via
\begin{equation}
\left( \begin{array}{c}
        Q_1 \\
        Q_2 \\
        Q_3 \end{array} \right)
\equiv {\bf Q} = {\cal U} {\bf P}.
\end{equation}
The transformation matrix ${\cal U}$ and its inverse ${\cal
U}^{-1}$ diagonalise the matrix ${\cal K}$ in Eq.\ (\ref{pqke}) by
definition,
\begin{equation}
{\cal K}_d \equiv \text{diag}(k_1,\, k_2,\, k_3) = {\cal U K
U}^{-1},
\end{equation}
where the eigenvalues $k_1$, $k_2$ and $k_3$ are roots of the
cubic characteristic equation
\begin{equation}
\label{characteristic}
k^3 + 2 D k^2 + (D^2 + \lambda^2 + \beta^2)
k + \beta^2 D = 0,
\end{equation}
the discriminant of which is identically
\begin{equation}
\label{discriminant} \Delta =  4 \beta^6 - \beta^4 D^2 + 12
\beta^4 \lambda^2 - 20 \beta^2 D^2 \lambda^2 + 4 D^4 \lambda^2 +
12 \beta^2 \lambda^4 + 8 D^2 \lambda^4 + 4 \lambda^6.
\end{equation}
Under most circumstances (for example, the individual cases of
$|\lambda| \gg D$ and $|\beta| \gg D$), the inequality $\Delta
> 0$ holds such that two of the three eigenvalues occur
predominantly as a complex conjugate pair, which may be
conveniently parameterised as
\begin{equation}
\label{k1k2} k_{1,\,2} =  -d \pm i \omega,
\end{equation}
where $d$ and $\omega$, both real and positive, are readily
interpreted as the {\it effective damping factor} and {\it
effective oscillation frequency} respectively.  These
phenomenological  parameters arise since the QKEs couple the
damped and oscillatory aspects of the time evolution, and are to
be compared with what could be called the {\it bare damping
factor} $D$ and the {\it bare matter-affected oscillation
frequency} $\Omega \equiv \sqrt{\beta^2 + \lambda^2}$. The
remaining root $k_3$ of Eq.\ (\ref{characteristic}) is real and
negative, and bears several simple but elucidating relationships
to $k_1$ and $k_2$,
\begin{eqnarray}
k_3 &=& -\frac{\beta^2 D}{d^2 + \omega^2}, \label{k3} \\ d & = &
D+\frac{k_3}{2}, \label{d} \\ \omega^2 &=& \lambda^2 + \beta^2 +
k_3 D + \frac{3}{4} k_3^2, \label{omega}
\end{eqnarray}
to be further discussed in Section \ref{classical}. These equations show
that $k_3$ quantifies the ``misalignment'' between the effective and
bare damping factor, and between the effective and bare oscillation frequency.

The only (so far) unambiguously identified instance for which the
discriminant is zero or negative ($\Delta \leq 0$) occurs when the
conditions $D \gg |\beta|$ and $\lambda^2 \stackrel{<}{\sim}
\frac{\beta^4}{4 D^2}$ are simultaneously satisfied, in which case
all eigenvalues are real and generally distinct with the exception
of $\Delta =0$, where at least two roots are
equal.\footnote{Equations (\ref{k1k2}) to (\ref{omega}) are
equally valid for cases where $\Delta \leq 0$ if we extend the
definition of $\omega$ to include imaginary values, although
$\omega$'s physical meaning is then, of course, different.} The
interested reader is referred to Appendix \ref{cubic} for the
mathematical details.  For clarity, we shall deal exclusively with
off-resonance evolution in this subsection, which translates
loosely into requiring $|\lambda| > D,\, |\beta|$, so as to ensure
the existence of complex conjugate eigenvalues. The study of
resonance behaviour i.e., where $\lambda \simeq 0$, is deferred to
Sec.\ \ref{zerolambda}.

For $\lambda \neq 0$, the matrix ${\cal U}^{-1}$ consists of the
normalised eigenvectors
\begin{equation}
\label{u-1matrix} \kappa_i = \frac{1}{{\cal N}_i} \left(
\begin{array}{c}
                    1 \\
                    -\frac{D+k_i}{\lambda} \\
                    -\beta \frac{D+k_i}{\lambda k_i} \end{array}
                    \right),
\end{equation}
in the columns, while the row vectors
\begin{equation}
\label{umatrix}
v_i = - {\cal N}_i k_i \left(  \frac{1}{D}
\prod_{j \neq i}
                    \frac{D+k_j}{k_i-k_j}, \quad
                    \lambda \prod_{j \neq i} \frac{1}{k_i-k_j},
                    \quad
                    \frac{\lambda}{\beta D} \prod_{j \neq i}
                    \frac{k_j}{k_i-k_j} \right),
\end{equation}
constitute the inverse matrix ${\cal U}$, with the $i$th
normalisation factor given by\footnote{A minor issue pertains to
the definition of the normalisation factor --- to be clarified
later where relevant.}
\begin{equation}
\label{normalisation} {\cal N}_i=\frac{1}{|\lambda k_i|}
\sqrt{\lambda^2 |k_i|^2 + (\beta^2 + |k_i|^2) |D+k_i|^2},
\end{equation}
or explicitly, with the aid of Eqs.\ (\ref{k3}) to (\ref{omega}),
\begin{eqnarray}
{\cal N}_1 &=& {\cal N}_2 = \frac{1}{|\lambda|} \sqrt{2 [\lambda^2
+ \beta^2 + k_3 (D + k_3)]},\nonumber \\ {\cal N}_3 &=&
\frac{\beta}{|\lambda k_3|} \sqrt{D(D+k_3)}.
\end{eqnarray}
In this new basis ${\cal S}_q$, the unit vector $\hat{q}_3$
represents the axis about which ${\bf P}$ precesses. This
precession axis coincides, by definition, with the matter
potential vector ${\bf V}$ in the absence of collisions (i.e.,
$D=0$), residing entirely on the $xz$-plane when viewed in the
fixed coordinate system ${\cal S}$. Otherwise the alignment is
inexact in the general case, where a nonzero D generically endows
$\hat{q}_3$ with a small $y$-component evident in the
transformation matrix ${\cal U}^{-1}$ in Eq.\ (\ref{u-1matrix}),
that is,
\begin{equation}
\hat{q}_3 = {\cal U}^{-1}_{x3} \hat{x} + {\cal U}^{-1}_{y3}
\hat{y} + {\cal U}^{-1}_{z3} \hat{z},
\end{equation}
where
\begin{eqnarray}
\label{q3components}
\label{q3} \left|{\cal U}^{-1}_{x3} \right|^2
&=& \frac{\beta^2}{(D+k_3)^2 + \lambda^2 + \beta^2} \left[ 1 -
\frac{D(D+k_3)}{(D+k_3)^2 + \lambda^2 + \beta^2} \right]
\stackrel{D = 0}{\longrightarrow} \frac{\beta^2}{\lambda^2 +
\beta^2},\ \nonumber
\\ \left|{\cal U}_{y3}^{-1}\right|^2 &=& \frac{\beta^2
D (D+k_3)}{\left[(D+k_3)^2 + \lambda^2 +
\beta^2\right]^2}\stackrel{D =0}{\longrightarrow} 0,\nonumber \\
\left|{\cal U}^{-1}_{z3} \right|^2 &=& \frac{(D+k_3)^2 + \lambda^2
}{(D+k_3)^2 + \lambda^2 + \beta^2} \stackrel{D =
0}{\longrightarrow} \frac{\lambda^2}{\lambda^2 + \beta^2}.
\end{eqnarray}
The $D=0$ (i.e., $\hat{q}_3
\parallel {\bf V}$) limit of Eq.\ (\ref{q3}) is to be compared
with Eq.\ (\ref{potential}). The variables $Q_1$ and $Q_2$
quantify the actual precession in conjunction with their
associating complex vectors, $\hat{q}_1$ and $\hat{q}_2$, that
sweep the plane perpendicular to $\hat{q}_3$.\footnote{This
description of the ${\cal S}_q$ coordinate system is actually
technically inaccurate since the instantaneous diagonal basis is
not orthogonal, and $\hat{q}_{1,2}$ are complex. However, as a
visual aid, it is more than adequate.} Figure \ref{basis} shows
schematically the relationship between the ${\cal S}$ and ${\cal
S}_q$ bases.

Rewriting Eq.\ (\ref{pqke}) in the new instantaneous ${\cal S}_q$
basis, we obtain
\begin{equation}
\label{qDE}
\frac{\partial {\bf Q}}{\partial t} = {\cal K}_d {\bf
Q} - {\cal U} \frac{\partial {\cal U}}{\partial t}^{-1} {\bf Q},
\end{equation}
where the $3 \times 3$ matrix ${\cal U} \frac{\partial {\cal
U}}{\partial t}^{-1}$ exemplifies the system's explicit dependence
on $\frac{\partial D}{\partial t}$, $\frac{\partial
\lambda}{\partial t}$ and $\frac{\partial \beta}{\partial t}$, and
in general contains nonzero off-diagonal entries.  Under certain
circumstances, these time derivatives are inconsequentially small
relative to terms in the diagonal matrix ${\cal K}_d$ (see Ref.\
\cite{bvw} for the relevant constraints).  When these bounds hold,
we are entitled to take the adiabatic limit, defined by ${\cal U}
\frac{\partial {\cal U}}{\partial t}^{-1} \simeq 0$, or
equivalently,
\begin{equation}
{\cal U}(D+\delta D,\, \lambda + \delta \lambda,\, \beta + \delta
\beta) {\cal U}^{-1}(D,\, \lambda,\, \beta) \simeq
\text{diag}(1,1,1),
\end{equation}
such that the basis ${\cal S}_q(D,\,\lambda,\, \beta)$ maps
directly onto the neighbouring ${\cal S}_q(D+\delta
D,\,\lambda+\delta \lambda,\,\beta + \delta \beta)$ as the
parameters $D$, $\lambda$ and $\beta$ slowly vary with time.

The formal
solution to the now completely decoupled system of differential
equations for ${\bf Q}$ is simply
\begin{equation}
\label{qsolution}
{\bf Q}(t) \simeq \text{diag} \left(e^{\int^t_0
k_1(t') dt'},\; e^{\int^t_0 k_2(t') dt'},\;
    e^{\int^t_0 k_3(t') dt'} \right)
                                {\bf Q}(0),
\end{equation}
and consequently,
\begin{equation}
\label{psolution} {\bf P}(t) \simeq {\cal U}^{-1}(t) \text{diag}
\left(e^{\int^t_0 k_1(t') dt'},\; e^{\int^t_0 k_2(t') dt'},\;
e^{\int^t_0 k_3(t') dt'} \right) {\cal U}(0) {\bf P}(0).
\end{equation}
Note that this solution is ``formal'' because the eigenvalues
$k_i$ depend on the asymmetry $L_{\alpha}$, and hence on $P_z$,
through the function $\lambda$.

At this point, we alert the reader to a mathematical subtlety.
Equations (\ref{qsolution}) and therefore (\ref{psolution}) are in
fact poorly defined if the time integration encompasses regions
where two or more eigenvalues are identical.  In this case, the
matrix ${\cal K}$ has momentarily less than three distinct
eigenvectors, thereby rendering the transformation matrix ${\cal
U}^{-1}$ uninvertible.  Fortunately, such instances, besides their
rarity (see Appendix \ref{cubic}), have virtually no bearings on
the outcome as we shall see in due course.

From Eq.\ (\ref{psolution}), we extract a formal expression for $P_y(t)$
in terms of $P_z(t)$,
\begin{equation}
\label{pypz} P_y(t) \simeq \frac{{\cal U}^{-1}_{y1}(t)e^{\int^t_0
k_1(t') dt'} Q_1(0) + {\cal U}^{-1}_{y2}(t)e^{\int^t_0 k_2(t')
dt'} Q_2(0) + {\cal U}^{-1}_{y3}(t)e^{\int^t_0 k_3(t') dt'}
Q_3(0)} {{\cal U}^{-1}_{z1}(t)e^{\int^t_0 k_1(t') dt'} Q_1(0) +
{\cal U}^{-1}_{z2}(t)e^{\int^t_0 k_2(t') dt'} Q_2(0) + {\cal
U}^{-1}_{z3}(t)e^{\int^t_0 k_3(t') dt'} Q_3(0)} P_z(t).
\end{equation}
Recall from Eq.\ (\ref{k1k2}) that the eigenvalues $k_1$ and $k_2$
are prevailingly composites of a real damping factor $d$ and an
imaginary oscillatory component $\omega$.  Given an ample
$\lambda$ such that the effective oscillation frequency reads
$\omega^2 \simeq \lambda^2$ in Eq.\ (\ref{omega}), the real
eigenvalue $k_3$ is guaranteed to be small according to Eq.\
(\ref{k3}), while the phenomenological damping becomes $d \simeq
D$ by Eq.\ (\ref{d}). Since $k_3$ scales with $D$, a comparatively
large $d$ will always quickly reduce the corresponding
exponentials in Eq.\ (\ref{pypz}) to zero relative to the
``decay'' time of their $k_3$ counterpart, wiping out the
accompanying rapid oscillations in the process. The condition $D
\gg |\beta|$ is a bonus which contributes to accomplishing the
said exponential damping at an even faster rate over the time
scale of the $k_3$ decay. Installing the {\it collision dominance}
approximation\footnote{The resulting approximate evolution equations
actually involve the joint action of collisions and oscillations.
However, we adopt the phrase ``collision dominance'' because of
the crucial role played by damping.}
\begin{equation}
\label{exp}
e^{\int^t_0 k_{1,\,2}(t') dt'} \to 0,
\end{equation}
in Eq.\ (\ref{pypz}), we obtain
\begin{equation}
\label{pypzapprox}
P_y(t) \simeq \frac{{\cal
U}_{y3}^{-1}(t)}{{\cal U}_{z3}^{-1}(t)}P_z(t) = \frac{k_3}{\beta}
P_z(t),
\end{equation}
which, together with its antineutrino analogue, allows us to
express the neutrino asymmetry evolution equation  in Eq.\
(\ref{exactdldt}) in an approximate form,
\begin{equation}
\label{approxdldt} \frac{dL_{\alpha}}{dt} \simeq \frac{1}{2
n_{\gamma}} \int \left\{ k_3 [N_{\alpha}(p)-N_s(p)] -
\overline{k}_3 [\overline{N}_{\alpha}(p)-\overline{N}_s(p)]
\right\} dp.
\end{equation}
Observe that Eq.\ (\ref{approxdldt}) involves {\it only} the
$\nu_{\alpha}$ and $\nu_s$ distribution functions at any given
time. Interestingly, phase dependence has been eliminated by the
adiabatic procedure in conjunction with appreciable damping
(collision dominance); the coherence status of the neutrino
ensemble has minimal influence the asymmetry's time evolution.
Geometrically, the quantity $P_y(t)/P_z(t)$ in Eq.\
(\ref{pypzapprox}) is but the instantaneous ratio of the $y$- and
$z$-components of the axis $\hat{q}_3$ in the limit of
zero-amplitude precession. We shall henceforth refer to this
condition as the {\it Boltzmann limit} of the QKEs. Note also that
Eq.\ (\ref{approxdldt}) has the form of a classical-style rate
equation [see Eq.\ (\ref{rateeqn}) below] with the neutrino and
antineutrino transition rates computed to be $|k_3|/2$ and
$|\overline{k}_3|/2$ respectively.

For computational purposes, an auxiliary expression describing
sterile neutrino production for each momentum state $p$ may be
obtained similarly in the Boltzmann limit \cite{bvw},
\begin{equation}
\label{sterile} \frac{d}{dt} \left[
\frac{N_s(p)}{N^{\text{eq}}(p,0)} \right] = - \frac{1}{2} \beta
P_y(p)  \simeq  -\frac{k_3}{2}  \left[ \frac{N_{\alpha} (p) -
N_s(p)}{N^{\text{eq}}(p,0)} \right],
\end{equation}
to be employed in tracking the quantity $N_s(p)$ in Eq.\
(\ref{approxdldt}). The $\nu_{\alpha}$ distribution function in
the same equation is approximated to be
\begin{equation}
\label{instant} N_{\alpha}(p) \simeq N^{\text{eq}}(p,\mu),
\end{equation}
with the interpretation that the neutrino momentum state $p$ is
instantaneously repopulated.  Thus Eqs.\ (\ref{approxdldt}) to
(\ref{instant}) form a fully serviceable set to be simultaneously
solved to give $L_{\alpha}$ as a function of time.

\subsection{Resonance behaviour: the $\lambda \simeq 0$ case}
\label{zerolambda}

Reference \cite{bvw} did not provide a full analysis of
collision-affected adiabatic evolution through an MSW resonance.
Since resonance behaviour is a very important issue in the study
of lepton asymmetry growth, we now provide a careful treatment of
this topic.

As an incentive to study the vicinity of an MSW resonance, let us
observe the solutions $k$ to the characteristic equation, Eq.\
(\ref{characteristic}), at exactly $\lambda=0$,
\begin{equation}
\left. k \right|_{\lambda=0}=-D,\quad - \frac{D}{2} \pm
\frac{\sqrt{D^2 - 4 \beta^2}}{2}.
\end{equation}
Evidently, whether the square root evaluates to an imaginary or a
real number is conditional to the relative sizes of the arguments
$D$ and $\beta$.  If the former ensues, we are forced to identify
the root $-D$ with the real eigenvalue $k_3$ which is now twice
the magnitude of the effective damping factor $d= D/2$ such that the
off-resonance ``damping versus decay'' rationale of Sec.\
\ref{boltzmannlimit} no longer applies. The latter scenario is,
contrary to off-resonance behaviour, non-oscillatory, and requires
a new interpretation.

\subsubsection{Case 1: $|\beta| \stackrel{>}{\sim} D$}

There are two distinct situations covered under this heading. The
first is where adiabatic and collision dominated evolution occurs
on either side of the resonance, with the condition $|\beta|
\stackrel{>}{\sim} D$ maintained during the crossing.  This
amounts to requiring that
\begin{eqnarray}
\label{crap}
\sin^2 2 \theta &\stackrel{>}{\sim}& 3.6 \times
10^{-24} y_{\alpha}^2 \frac{|\Delta m^2| y}{L^{(\alpha)3}}
\nonumber \\ &\sim & 10^{-7},
\end{eqnarray}
where $y = p/T$ is a dimensionless quantity not to be confused
with $y_{\alpha}$, and we have used Eqs.\ (\ref{betalambda}) and
(\ref{dp}) evaluated at resonance, $\cos 2 \theta - a(p) \simeq
0$. The last approximate inequality in Eq.\ (\ref{crap}) is
established by setting $y \simeq 3.15$ (i.e., average momentum),
$|\Delta m^2| \sim 1 \, \text{eV}^2$, and $L^{(\alpha)} \sim
10^{-5}$ is a typical value for the asymmetry immediately after
exponential growth.

The second scenario is where $D$ is genuinely small enough so that
the collision dominance approximation cannot be made even outside
of the resonance region. This situation typically obtains,
independently of the vacuum oscillation parameters, at lower
temperatures since $D$ decreases with temperature as $T^5$
according to Eq.\ (\ref{dp}).

It is easy to convince oneself by inspecting the discriminant in
Eq.\ (\ref{discriminant}) that a dominating $\beta$ ensures the
existence of complex eigenvalues, and thereby preserves the
precessive nature of the evolution of the polarisation vector
${\bf P}$ for all $\lambda$. At $\lambda = 0$, the solutions to
the characteristic equation, Eq.\ (\ref{characteristic}), are
\begin{eqnarray}
\label{zerolambdaevalues2} \left. k_{1,\,2} \right|_{\lambda=0,\,
|\beta| \stackrel{>}{\sim} D} &=& -\frac{D}{2} \pm i\frac{\sqrt{4
\beta^2-D^2}}{2}, \nonumber \\ \left. k_3 \right|_{\lambda=0,\,
|\beta| \stackrel{>}{\sim} D} &=& -D,
\end{eqnarray}
where the sole real root is always identified as $k_3$, which, as
mentioned earlier, is clearly larger than the real components of
$k_1$ and $k_2$, i.e., the effective damping factor $d = D/2$. The
corresponding transformation matrices evaluate to
\begin{eqnarray}
\label{superstrong}
\left. {\cal U}^{-1} \right|_{\lambda=0,\,
|\beta| \stackrel{>}{\sim} D}
    &=& \left(  \begin{array}{ccc}
                    0 & 0 & 1 \\
                    -\frac{1}{\sqrt{2}}\frac{k_1}{\beta}
                    & -\frac{1}{\sqrt{2}}\frac{k_2}{\beta}& 0 \\
                    -\frac{1}{\sqrt{2}} & -\frac{1}{\sqrt{2}}& 0
                    \end{array} \right), \nonumber \\ \left. {\cal U}
\right|_{\lambda=0,\, |\beta| \stackrel{>}{\sim} D}
    & = &   \left(  \begin{array}{ccc}
                    0 & i \beta \sqrt{\frac{2}{4 \beta^2-D^2}} &
                        -i k_2\sqrt{\frac{2}{4 \beta^2-D^2}}  \\
                    0 & -i \beta\sqrt{\frac{2}{4 \beta^2-D^2}} &
                        i k_1\sqrt{\frac{2}{4 \beta^2-D^2}}  \\
                    1 & 0 & 0 \end{array} \right),
\end{eqnarray}
where the precession axis $\hat{q}_3$ now lies
in the $x$-direction, and the vectors $\hat{q}_1$ and $\hat{q}_2$
trace out accordingly a surface roughly parallel to the
$yz$-plane.

Let us now track the evolution of ${\bf P}$, as depicted in Fig.\
\ref{smalldq3}, for the situation where collision dominance holds
on either side of the resonance crossing. As the matter potential
vector ${\bf V}$, initially at $(+\beta,\,0,\, +\lambda)$, is
rotated on the $xz$-plane and decoherence function $D$
independently modified, the instantaneous unit vector $\hat{q}_3$
moves through a continuum of predefined positions in the
$(+x,\,-y,\,+z)$ block dictated by the transformation matrix
${\cal U}^{-1}$, while ${\bf P}$ precesses about it.  The vector
$\hat{q}_3$ gradually aligns with the $x$-axis as we approach
$\lambda=0$, such that the decay of $Q_3$ becomes increasingly
dependent on the parameters that govern the evolution of $P_x$.
This is manifested in the escalating magnitude of the eigenvalue
$k_3$, which reaches a zenith of $|-D|$ when $\hat{q}_3$ and
$\hat{x}$ are completely parallel at $\lambda = 0$ by Eqs.\
(\ref{zerolambdaevalues2}) and (\ref{superstrong}), and $Q_3$ and
$P_x$ are momentarily equivalent.\footnote{A more insightful
interpretation of $k_3$ will be provided later in Sec.\
\ref{classical}.} Meanwhile, the two variables of interest, $P_y$
and $P_z$, become progressively more oscillatory as we cross the
resonance region for the following reason.  

In the adiabatic limit,
the  precessive behaviour of ${\bf P}$ is parameterised by
the complex conjugate variables
$Q_1$ and $Q_2$, and driven by 
the imaginary 
component of the corresponding eigenvalues $k_1$ and $k_2$.  
The magnitudes of the oscillatory components in $P_{x,y,z}$ 
are therefore dependent on two factors:  
(i) exponential damping of $Q_1$ and $Q_2$ 
through the real part of $k_1$ and $k_2$, and  
(ii) the ``intrinsic'' oscillation 
amplitudes prescribed by the transformation matrix ${\cal U}^{-1}$, or
equivalently, the projections of the instantaneous eigenvectors 
$\hat{q_1}$ and $\hat{q_2}$ onto the fixed ${\cal S}$ basis.
For the $|\beta| \stackrel{>}{\sim} D$ case, 
Eq.\ (\ref{zerolambdaevalues2}) shows that
the effective damping factor $d$ in  $k_1$ and $k_2$
  diminishes from $\sim D$ to an
  all-time low of $D/2$. 
   This means that the variables $Q_1$ and $Q_2$ are no longer preferentially
exponentially    
damped relative to the decay of $Q_3$ which now occurs at
   a comparable rate $\sim D$.   On this account alone, no single
   term in Eq.\ (\ref{pypz}) stands out as the presiding factor in the
   resonance region.  At the same time,
   the evolution of $P_y$ and $P_z$ is now almost totally
  described by $Q_1$ and $Q_2$,
  because of a growing alignment between the $q_1 q_2$- and
$yz$-planes.  Complete
alignment (and therefore purely oscillatory behaviour in $P_y$ and
$P_z$) is attained at exactly $\lambda=0$ as shown in Eq.\
(\ref{superstrong}).

On the other side of the resonance, $\hat{q}_3$ resides in the
$(+x,\,+y,\,-z)$ block while ${\bf V}=(+\beta,\, 0, -\lambda)$
continues to move down the $xz$-plane (see Fig.\ \ref{smalldq3}).
The variable $Q_3$ regains control  as $|\lambda|$ increases, and
the system advances in time as discussed in Sec.\
\ref{boltzmannlimit}.

The oscillatory terms at resonance play a pivotal role in
partially propelling the asymmetry growth through the MSW effect
(dominant mode for $|\beta| \gg D$, i.e., negligible collision
rate).  In fact, in regimes where collisions are completely absent
(i.e., $D=0$), one may recover from the approximate solution to
the QKEs in Eq.\ (\ref{psolution}) the common expression for the
adiabatic MSW survival probability \cite{bvw}.  Geometrically,
adiabatic $|\beta| \stackrel{>}{\sim} D$ evolution carries the
polarisation vector ${\bf P}$ from the $(+x,\, -y, \,+z)$ to the
$(+x,\, +y,\, -z)$ block across an MSW resonance following the
precession axis $\hat{q}_3$. But as far as Eq.\ (\ref{approxdldt})
is concerned, coherent MSW transitions are evidently dependent on
the history of the neutrino ensemble, and therefore outside the
Boltzmann limit. The approximate neutrino asymmetry evolution
equation, Eq.\ (\ref{approxdldt}), thereby collapses in the
proximity of a resonance under the condition $|\beta|
\stackrel{>}{\sim} D$.

\subsubsection{Case 2: $D \gg |\beta|$}
\label{dggbeta}

The case of $D \gg |\beta|$ exhibits vastly contrasting resonance
behaviour to the previously considered scenario. A genuinely large
$d \simeq D$ in denominator of $k_3$, that is,
\begin{equation}
\label{bigdk3} \left.k_3\right|_{D \gg \beta} \simeq -
\frac{\beta^2 D}{D^2+\lambda^2} + {\cal O}(\beta^4),
\end{equation}
according to Eqs.\ (\ref{k3}) to (\ref{omega}), will ensure the
latter's smallness even as $\lambda$ approaches the resonance
region.

An immediate consequence is a growing projection of
$\hat{q}_3$, initially in the $(+x,\, -y,\, +z)$ block, onto the
$yz$-plane, accompanied by an $x$-component that tends to zero
with a decreasing $\lambda$ in Eq.\ (\ref{q3components}).
Precession about $\hat{q}_3$ is sustained and quantified by the
complex functions $Q_1$ and $Q_2$ until we reach a domain where
the imaginary parts of the complex eigenvalues totally vanish. We
locate this territory by evaluating the discriminant of the
characteristic equation in Eq.\ (\ref{discriminant}) with the
condition $D \gg |\beta|$ in place (see Appendix \ref{cubic}). The
patch turns out to be minuscule, with an upper bound
\begin{equation}
\label{recipe} \lambda^2 \simeq \frac{\beta^4}{4D^2},
\end{equation}
to be compared with the nominal resonance width $|\Delta \lambda|
\simeq |\beta|$.  Here, the eigenvalues are all real and negative,
with their relative sizes being the only distinguishing feature.
Thus, in principle, any one root may be labelled $k_3$ and the
remaining two will automatically satisfy Eqs.\ (\ref{k1k2}) to
(\ref{omega}) with an imaginary $\omega$.  We restrict the choice
by matching the eigenvalues at the ``complex--real'' interface,
such that the smallest root is always designated
$k_3$.\footnote{Once matched, the root $k_3$ will remain smallest
in the ``real'' region since real eigenvalues must never cross
therein (except at the boundary) in the limit $D \gg |\beta|$, as
dictated by a nonzero discriminant in Eq.\ (\ref{discriminant}).}
The others, $k_1$ and $k_2$, are dissimilar but concurrently
large, thereby providing continued justification for Eq.\
(\ref{exp}) and consequently Eq.\ (\ref{pypzapprox}) in this
regime.  As an illustration, we calculate the eigenvalues at
$\lambda = 0$,
\begin{eqnarray}
\label{zerolambdaevalues} \left.k_1\right|_{\lambda=0,\,D \gg
|\beta|} &=& -D, \nonumber \\ \left. k_2\right|_{\lambda=0,\,D \gg
|\beta|} &=& -\frac{D}{2} - \frac{\sqrt{D^2-4\beta^2}}{2} \simeq
-D + {\cal O}(\beta^2), \nonumber \\ \left.k_3\right|_{\lambda =
0,\,D \gg |\beta|} &=& -\frac{D}{2} + \frac{\sqrt{D^2 - 4
\beta^2}}{2} \simeq -\frac{\beta^2}{D} + {\cal O}(\beta^4).
\end{eqnarray}
Note that the expression for $k_3$ in fact agrees very well with
Eq.\ (\ref{bigdk3}).

The unit vectors $\hat{q}_1$, $\hat{q}_2$
and $\hat{q}_3$ are similarly matched at the boundary.
Transformations between the fixed ${\cal S}$ and instantaneous
${\cal S}_q$ bases are as defined previously in Eqs.\
(\ref{u-1matrix}) and (\ref{umatrix}), although minimal physical
significance may be ascribed to the instantaneous basis beyond the
textbook description of a $3$-space spanned by three real linearly
independent vectors, where $\hat{q}_3$ is, incidentally, most
aligned with $\hat{z}$.  The matrices ${\cal U}^{-1}$ and ${\cal
U}$ evaluated at $\lambda =0$ are routinely displayed below,
\begin{eqnarray}
\label{affirmation}
\left. {\cal U}^{-1} \right|_{\lambda=0,\,D
\gg |\beta|}
    &=& \left(  \begin{array}{ccc}
                    1 & 0 & 0 \\
                    0 &\frac{k_2}{\sqrt{\beta^2 +k_2^2}}
                    & \frac{k_3}{\sqrt{\beta^2 + k_3^2}} \\
                    0&\frac{\beta}{\sqrt{\beta^2 + k_2^2}}
                    & \frac{\beta}{\sqrt{\beta^2 + k_3^2}}
                    \end{array} \right), \nonumber \\ \left. {\cal U}
\right|_{\lambda=0,\,D \gg |\beta|}
    & = &   \left(  \begin{array}{ccc}
                    1 & 0 & 0 \\
                    0 & \sqrt{\frac{\beta^2 + k_2^2}{D^2 - 4 \beta^2}} &
                        -\frac{k_3}{\beta}\sqrt{\frac{\beta^2 + k_2^2}
{D^2 - 4 \beta^2}}  \\
                    0 & -\sqrt{\frac{\beta^2 + k_3^2}{D^2 - 4 \beta^2}} &
                        \frac{k_2}{\beta}\sqrt{\frac{\beta^2 + k_3^2}
{D^2 - 4 \beta^2}}
                    \end{array} \right),
\end{eqnarray}
in which the quantities $k_2$ and $k_3$ are the relevant
eigenvalues listed in Eq.\ (\ref{zerolambdaevalues}), and the
ratio ${\cal U}^{-1}_{y3}/{\cal U}^{-1}_{z3} = k_3/\beta$ remains
unchanged from the $\lambda \neq 0$ case.  Hence the asymmetry
continues to evolve in the Boltzmann limit at resonance in the $D
\gg |\lambda|,\, |\beta|$ regime, as described previously by Eq.\
(\ref{approxdldt}).

Observe also in Eq.\ (\ref{affirmation}) that
the vector $\hat{q}_3$ is indeed entirely on the $yz$-plane at
$\lambda=0$. Thus the role of adiabatic evolution in the event of
$D \gg |\beta|$ is to bring the polarisation ${\bf P}$ from the
$(+x,\, -y, \, +z)$ to the $(-x,\, -y, +z)$ block as we rotate the
matter potential vector ${\bf V}$ from $(+\beta,\, 0, +\lambda)$
to $(+\beta,\, 0, -\lambda)$, as shown in Fig.\ \ref{bigdq3}.
This is in total contrast with the $|\beta| \stackrel{>}{\sim}D$
and, indeed, the $D=0$ cases, and entails a new definition for the
${\cal U}^{-1}$ and ${\cal U}$ normalisation factor to maintain
the mathematical correctness of Eqs.\ (\ref{qsolution}) and
(\ref{psolution}), that is,
\begin{equation}
\left.{\cal N}_i\right|_{D \gg |\beta|}=\frac{1}{\lambda |k_i|}
\sqrt{\lambda^2 |k_i|^2 + (\beta^2 + |k_i|^2) |D+k_i|^2}.
\end{equation}
Compared with Eq.\ (\ref{normalisation}), this new definition
guarantees that the $x$-component of $\hat{q}_3$, ${\cal
U}^{-1}_{x3}$, (and not ${\cal U}^{-1}_{y3}$ and ${\cal
U}^{-1}_{z3}$, as in the $|\beta| \stackrel{>}{\sim}D$ case) flips
sign across an MSW resonance.

\subsubsection{Summary}

In this subsection, the following conclusions have been reached
pertaining to evolution through an MSW resonance:
\begin{enumerate}
\item Where the condition $D \gg |\beta|$ is satisfied,
asymmetry evolution (which
is just a special case of distribution function evolution) continues
to be well described by the Boltzmann limit. Sufficient damping
is in place to remove the coherence history of the ensemble as a
necessary and independent dynamical variable.
\item The case of $|\beta| \stackrel{>}{\sim} D$ displays completely
 different resonance behaviour since the damping versus decay rationale
supplied by $|\text{Re}(k_{1,2})| \gg |k_3|$ does {\it not} hold.
The MSW effect now dominates the dynamics of the system in a way
that cannot be described by a Boltzmann limit.
\end{enumerate}

\subsection{Static approximation}
\label{staticapprox}

We present here a brief account of the static approximation
\cite{longpaper,ftv}, partly to demonstrate that Eq.\
(\ref{approxdldt}) may be re-derived within this framework, but
also to introduce some terminology for later use. The derivation
given here is a development on that in Ref.\ \cite{longpaper}.

The static approximation
begins with the following observations \cite{longpaper,ftv}: There
are two means by which the lepton number of the universe may be
modified:  (i) oscillations between collisions, since matter
effects are dissimilar for neutrinos and antineutrinos,
 and (ii) the collisions
themselves which deplete $\nu_{\alpha}$ and
$\overline{\nu}_{\alpha}$ at different rates through physically
``measuring'' the respective $\nu_{\alpha} \leftrightarrow \nu_s$
and $\overline{\nu}_{\alpha} \leftrightarrow \overline{\nu}_s$
adiabatic matter-affected oscillation probabilities.

This vision of the role of collisions is essentially a
wave-function collapse hypothesis (projection postulate) that is
evidently closely related to the collisional decoherence
rigorously quantified through the $D$ function in the QKEs.
Exploitation of this picture has, in the past, led to the
acquisition of much physical insight (see Refs.\
\cite{longpaper,ftv}). There is a significant literature on the
relationship between the collapse hypothesis and systematic
treatments of quantal decoherence for open systems \cite{atomic}.
The general conclusion seems to be that the former often leads to
dynamical behaviour close to that derived from the formal ``master
equation'' or QKE approach. Using this type of picture for
active--sterile neutrino oscillations, one can derive a lepton
number evolution equation (static approximation) similar to that
obtained in the adiabatic limit of the QKEs in the collision
dominance regime \cite{longpaper,bvw}.

The approximation we shall discuss here neglects mechanism (i)
above. Mode (ii), generally predominant in the high
temperature regime, is
described by the rate equation
\begin{eqnarray}
\label{rateeqn} \frac{dL_{\alpha}}{dt} = \frac{1}{n_{\gamma}} \int
&& \left[- \Gamma (\nu_{\alpha} \to \nu_s,\, p) N_{\alpha}(p) +
\Gamma (\nu_s \to \nu_{\alpha},\, p) N_s(p) \right. \nonumber \\
&&+ \left. \Gamma (\overline{\nu}_{\alpha} \to \overline{\nu}_s,\,
p)\overline{N}_{\alpha}(p) -  \Gamma (\overline{\nu}_s \to
\overline{\nu}_{\alpha},\, p)\overline{N}_s(p) \right] dp.
\end{eqnarray}
The reaction rate
\begin{equation}
\Gamma (\nu_{\alpha} \to \nu_s,\,p) = \frac{\Gamma(p)}{2} \langle
P(\nu_{\alpha} \to \nu_s,\,p) \rangle,
\end{equation}
is the product of {\it half} of the collision rate $\Gamma(p)$,
and the probability that an initially active neutrino at time $t'$
would collapse to the sterile eigenstate at the time of collision,
$t$, averaged over the neutrino ensemble.\footnote{The factor of
$1/2$ arises from  the fact that $D$ equals $\Gamma/2$ rather than
$\Gamma$ in the QKEs. The discovery paper \cite{ftv} omitted this
factor.}
  In the adiabatic limit, the latter is given by
\begin{eqnarray}
\label{reactionrate} \langle P(\nu_{\alpha} \to \nu_s,\,p) \rangle
&\simeq& \frac{1}{2} - \frac{1}{2} \left\langle
\frac{\lambda(t)}{\Omega(t)} \frac{\lambda (t')}{\Omega(t')} +
\frac{\beta(t)}{\Omega(t)} \frac{\beta(t')}{\Omega(t')}  \cos
\left[ \int^t_{t'} \Omega(t'') dt'' \right]
\right\rangle_{\text{ens}} \nonumber \\ & = & \frac{1}{2} -
\frac{1}{2} \left\langle c 2 \theta_m(t) c 2 \theta_m(t') + s 2
\theta_m(t) s 2 \theta_m(t') \cos \left[ \int^t_{t'} \Omega(t'')
dt'' \right] \right\rangle_{\text{ens}},
\end{eqnarray}
where $\Omega = \sqrt{\lambda^2 + \beta^2}$ is the matter-affected
oscillation frequency, $c 2\theta_m = \cos 2 \theta_m$ and $s
2\theta_m = \sin 2 \theta_m$ quantify the matter-affected mixing
angle, the subscript ``ens'' denotes a type of ensemble average,
and all quantities are  functions of momentum $p$. It follows
 from Eq.\ (\ref{reactionrate}) that $\Gamma(\nu_{\alpha}
\to \nu_s,\, p) = \Gamma(\nu_s \to \nu_{\alpha},\, p)$, and
likewise for their antineutrino counterparts.

The ensemble average phase $\langle \cdots \rangle_{\text{ens}}$
 is computed assuming that
collisions disrupt coherent evolution by resetting the phase term
 in the brackets to zero.  Consider the
$\nu_{\alpha}$ ensemble at time $t$.  The fraction that has
survived resetting since time $t'$ is postulated to be
\begin{equation}
z(t,\, t') = \exp\left[- \int_{t'}^{t} D(t'') dt'' \right].
\end{equation}
Thus the portion of neutrinos that is reset between $t'$ and $t'+
dt'$ and then propagates freely to time $t$ without further
encounters is simply
\begin{equation}
d z(t,\, t') = D(t') \exp \left[- \int_{t'}^{t} D(t'') dt''\right]
dt',
\end{equation}
contributing the phase $c 2 \theta_m(t) c 2 \theta_m(t') + s 2
\theta_m(t) s 2 \theta_m(t') \cos \left[ \int^t_{t'} \Omega(t'')
dt'' \right]$ to the ensemble at time $t$.
The {\it ensemble average phase}
is a weighted sum of these contributions originating in a time
interval extending from $t'=t_i$ in the past, to the present
$t'=t$:
\begin{eqnarray}
\label{ulysses} \lefteqn{\left\langle c 2 \theta_m(t) c 2
\theta_m(t') + s 2 \theta_m(t) s 2 \theta_m(t') \cos \left[
\int^t_{t'} \Omega(t'') dt'' \right] \right\rangle_{\text{ens}}}
\nonumber \\ &=& \int^{t'=t}_{t'=t_i}D(t') e^{-
\int_{t'}^{t}D(t'') dt''} \left\{ \frac{\lambda(t)}{\Omega(t)}
\frac{\lambda (t')}{\Omega(t')} + \frac{\beta(t)}{\Omega(t)}
\frac{\beta(t')}{\Omega(t')}  \cos \left[ \int^t_{t'} \Omega(t'')
dt'' \right] \right\} dt' \nonumber
\\ &=& \frac{\lambda(t)}{\Omega(t)} \int^t_{t_i} e^{- \int_{t'}^{t} D(t'')
dt''} \frac{D(t') \lambda(t')}{\Omega(t')} dt' + \frac{1}{2}
\frac{\beta(t)}{\Omega(t)} \left[ \int^t_{t_i}  e^{- \int_{t'}^{t}
(D + i \Omega) dt''}  \frac{D(t') \beta(t')}{\Omega(t')} dt' +
\text{c.c.} \right],
\end{eqnarray}
where c.c.\ denotes complex conjugate.  Integrating by parts, the
first integral  evaluates to
\begin{eqnarray}
\label{sticky} \lefteqn{\frac{\lambda(t)}{\Omega(t)} \int^t_{t_i}
D(t') e^{- \int_{t'}^{t} D(t'') dt''}
\frac{\lambda(t')}{\Omega(t')} dt'} \nonumber \\ &=&
\frac{\lambda^2(t)}{\Omega^2(t)} - e^{-\int^t_{t_i} D(t'') dt''}
\frac{\lambda(t)}{\Omega (t)} \frac{\lambda(t_i)}{\Omega(t_i)} -
\frac{\lambda(t)}{\Omega(t)} \int^t_{t_i} e^{-\int^t_{t'} D(t'')
dt''} \frac{d}{dt'} \left[\frac{ \lambda(t')}{\Omega(t')} \right]
dt' \nonumber \\ &\simeq& \frac{\lambda^2(t)}{\Omega^2(t)}.
\end{eqnarray}
The last approximate equality arises from the static approximation
which assumes negligible dependence on the time derivatives of $D$
and $\Omega$. The integration limit $t-t_i \to \infty$ has also
been utilised to eliminate the definite integral
$\exp[-\int^t_{t_i} D(t'') dt'']$ in Eq.\ (\ref{sticky}).

Applying the same procedure on the second term in Eq.\
(\ref{ulysses}), we obtain
\begin{eqnarray}
\label{tacky} \lefteqn{\frac{\beta(t)}{\Omega(t)} \int^t_{t_i} (D
+ i \Omega) e^{-\int^t_{t'} (D + i \Omega) dt''}
\frac{\beta(t')}{\Omega(t')} \frac{D(t')}{D(t') + i \Omega(t')}
dt'} \nonumber
\\ &=& \frac{\beta^2(t)}{\Omega^2(t)}\frac{D(t)}{D(t) + i
\Omega(t)}
 - e^{-\int^t_{t_i} (D + i \Omega) dt''}
 \frac{\beta(t)}{\Omega(t)} \frac{\beta(t_i)}{\Omega(t_i)}
 \frac{D(t_i)}{D(t_i) + i\Omega(t_i)} \nonumber \\
 && \hspace{3cm} - \frac{\beta(t)}{\Omega(t)}
 \int^t_{t_i} e^{-\int^t_{t'} (D + i \Omega) dt''} \frac{d}{dt'}
 \left[ \frac{\beta(t')}{\Omega(t')} \frac{D(t')}{D(t') + i
 \Omega(t')}  \right] dt'\nonumber \\
 &\simeq& \frac{\beta^2(t)}{\Omega^2(t)} \frac{D(t)}{D(t) + i
 \Omega(t)},
 \end{eqnarray}
and similarly for its complex conjugate.  Thus Eqs.\
(\ref{reactionrate}), (\ref{sticky}) and (\ref{tacky}) combine to
give
\begin{equation}
\langle P(\nu_{\alpha} \to \nu_s,\,p) \rangle \simeq \frac{1}{2} -
\frac{1}{2} \langle \cdots \rangle_{\text{ens}} \simeq \frac{1}{2}
\frac{\beta^2(t)}{D^2(t) + \Omega^2(t)},
\end{equation}
and the original rate equation in Eq.\ (\ref{rateeqn}) becomes
\begin{equation}
\frac{dL_{\alpha}}{dt} \simeq \frac{1}{2 n_{\gamma}} \int \left\{
k_3^{\text{static}} [N_{\alpha}(p)-N_s(p)] -
\bar{k}_3^{\text{static}}
[\overline{N}_{\alpha}(p)-\overline{N}_s(p)] \right\} dp,
\end{equation}
with
\begin{equation}
\label{static} k^{\text{static}}_3 = -\frac{\beta^2
D}{D^2+\lambda^2 + \beta^2},
\end{equation}
and a similarly defined antineutrino counterpart
$\bar{k}_3^{\text{static}}$ evaluated at time $t$.  Equation
(\ref{static}) is to be compared to the exact eigenvalue $k_3$
defined in Eqs.\ (\ref{k3}) to (\ref{omega}), and to its
approximate form in the limit $D,\, |\lambda| \gg |\beta|$
displayed in Eq.\ (\ref{bigdk3}). The exact $k_3$ agrees with
$k^{\text{static}}_3$ when the small $k_3$ contributions in Eqs.\
(\ref{d}) and (\ref{omega}) are neglected. This clarifies a point
of confusion raised at the beginning of Sec.\ II C in Ref.\
\cite{bvw} that was later resolved, for the first time, in Ref.\
\cite{keith}.

\subsection{Numerical results}

For the purpose of book-keeping,
we include here a report on the
results of numerically integrating Eqs.\ (\ref{approxdldt}) and
(\ref{sterile}) together with the approximation of instantaneous
repopulation given by Eq.\ (\ref{instant}). Calculations are
performed on several sets of $\nu_{\alpha} \leftrightarrow \nu_s$
oscillation parameters $\Delta m^2$ and $\sin^2 2 \theta_0$ for
two choices of $k_3$:
\begin{enumerate}
\item  Exact $k_3$. The real root $k_3$ of the characteristic
equation, Eq.\ (\ref{characteristic}), is computed by iterations
at each time step for all neutrino momentum bins. In the event of
three real solutions, we invariably choose the one smallest in
magnitude as justified in Sec.\ \ref{zerolambda}. This
calculation, codenamed \texttt{exactk3}, has not been previously
attempted.
\item Static $k_3$. In the \texttt{statick3} code, we adopt the
heuristically derived $k_3^{\text{static}}$ of Eq.\ (\ref{static})
which, in the past, has very successfully generated neutrino
asymmetry growths that closely mimic solutions of the exact QKEs
\cite{longpaper}. The present independent calculation incorporates
a distinct decoherence function $\overline{D}(p)$ for the
antineutrino ensemble that was formerly taken to be identical to
its neutrino counterpart.
\end{enumerate}
Results from \texttt{exactk3} and \texttt{statick3} for two
representative sets of $\Delta m^2$ and $\sin^2 2 \theta_0$
are shown in Figs.
\ref{asym2} and \ref{asym3}.  Solutions to the exact
QKEs (\texttt{qke}) for the same oscillation
parameters, computed independently for this work, are also
presented in the same figures for the purpose of comparison as
well as to demonstrate
that the QKEs are indeed numerically tractable.

In all three codes, time integration is achieved using
 the fifth order
Cash--Karp Runge--Kutta
method with an embedded fourth order formula for adaptive stepsize
control \cite{nr}. The
neutrino momentum distribution is discretised on a logarithmically
spaced mesh following Ref.\ \cite{p&r}, and a
summation over
all momentum bins is performed at
each time step.  For extra numerical
stability, the redundant Eq.\ (\ref{exactdldt}) is also built into
the \texttt{qke} code, to be simultaneously solved with Eq.\
(\ref{qkes}).  This feature essentially serves as a safety net that
insures against errors arising from
taking the difference between two large numbers.

It is clear from Fig.\ \ref{asym2} and \ref{asym3} that for all
practical purposes, \texttt{exactk3} and \texttt{statick3}
produce indistinguishable results (and we add in parentheses in conclusion
that the former code is time-consuming to execute).

\section{The classical oscillator}
\label{classical}

Much of the discussion in the preceding section, particularly on
the meaning of the eigenvalues, was conducted in jargons borrowed
freely from the damped simple harmonic oscillator.  We now present
a full classical analogue that models the individual time
development of the variables $P_x$, $P_y$ and $P_z$.

Consider a classical system exhibiting damped simple harmonic
motion about a midpoint $x_0$ that simultaneously ``decays'' with
time, such as portrayed in Fig.\ \ref{osc} and by the following
second and first order ordinary differential equations,
\begin{equation}
\label{2nd} \frac{d^2 (x - x_0)}{dt^2} + 2 \gamma \frac{d
(x-x_0)}{dt} + \nu_0^2 (x - x_0) = 0,
\end{equation}
and
\begin{equation}
\label{1st} \frac{d x_0}{dt} = - \xi x_0.
\end{equation}
The quantity $\xi$ is recognised as a decay constant, $\nu_0$ is
the natural frequency of the oscillator, and $\gamma$ is the
damping factor that couples to the oscillator's velocity to effect
a dissipative force.

The  variable $x$ may be solved for exactly
by inflating Eqs.\ (\ref{2nd}) and (\ref{1st}) conjointly to
produce a third order ordinary differential equation
\begin{equation}
\label{3rd} \frac{d^3x}{dt^3} + (2 \gamma + \xi)\frac{d^2x}{dt^2}
+ (2 \gamma \xi + \nu_0^2) \frac{dx}{dt} + \nu_0^2 \xi x = 0.
\end{equation}
Returning briefly to the QKEs (with negligible repopulation
function) in Eq.\ (\ref{pqke}), we observe that each component of
the polarisation vector ${\bf P}$ may alternatively be separately
solved by expanding the said system of homogeneous equations into
three mutually independent third order differential equations, one
per variable.  The inflated differential equations are
\begin{equation}
\label{thirdorder} \frac{\partial^3 {\bf P}}{\partial t^3} + 2 D
\frac{\partial^2 {\bf P}}{\partial t^2} + (D^2 + \lambda^2 +
\beta^2) \frac{\partial {\bf P}}{\partial t} + \beta^2 D {\bf P} =
0,
\end{equation}
for the case of time-independent parameters $D$, $\lambda$ and $\beta$.

Suppose $x$ is the parallel of $P_z$, i.e., the
normalised difference between
the $\nu_{\alpha}$ and $\nu_s$ distribution functions.\footnote{We shall
restrict our discussions to the evolution of $P_z$ since it is the
only variable that represents a tangible quantity, although the
analogy is equally applicable to $P_x$ and $P_y$.}  Comparison of
Eqs.\ (\ref{3rd}) and (\ref{thirdorder}) immediately leads to the
following identifications,
\begin{eqnarray}
\gamma & = & D-\frac{\xi}{2},\label{gamma}
\\ \nu_0^2 & = & (D - \xi)^2 + \lambda^2 + \beta^2,\label{nuzero}\\
\xi & = & \frac{\beta^2 D}{\nu_0^2}. \label{xi}
\end{eqnarray}
With some minor algebraic manipulation, the real quantity $\xi$
can be shown to satisfy the characteristic equation in Eq.\
(\ref{characteristic}).  Thus $\xi$ is identically $-k_3$, as the
reader would expect, and its role is to {\it equalise} the
neutrino distribution functions $N_{\alpha}$ and $N_s$, such as in
a reaction $A \rightleftharpoons B$ where the forward and reverse
reaction rate constants are both $\left|k_3\right|/2$.  From this
perspective, the Boltzmann limit simply consists of approximating
$P_z(t)$ as some midpoint $P_{z0}(t)$ [and similarly for
$P_y(t)$], which suffices if the oscillatory component $P_z(t) -
P_{z0}(t)$ is negligible relative to $P_{z0}(t)$ through
collision-induced damping and/or matter suppression.

Furthermore, suppose we wish to extend the definition of the
matter-affected mixing angle
\begin{equation}
\sin^2 2 \theta_m \equiv \frac{\beta^2}{\lambda^2 + \beta^2},
\end{equation}
to include collision effects by recognising that the denominator
of the above is simply the matter-affected oscillation frequency.
Then replacement with the natural frequency $\nu_0^2$ immediately
leads to an {\it effective matter and collision-affected mixing
parameter}
\begin{equation}
\label{Dmixing} \sin^2 2 \theta_{m,D} \equiv
\frac{\beta^2}{\nu_0^2},
\end{equation}
which, together with Eq.\ ({\ref{xi}), automatically grants $k_3$
a most intuitive interpretation; the reaction constant $k_3$
reflects on the individual neutrino state's ability to mix
(through $\sin^2 2 \theta_{m,D}$) and to collide (through $D$).
This is in complete agreement with heuristic derivations.  A quick
juxtaposition of the exact $k_3$ obtained from Eqs.\
(\ref{nuzero}) and (\ref{xi}),
\begin{equation}
k_3 = -\xi = - D \sin^2 2 \theta_{m,D} =
-\frac{\beta^2 D}{(D-\xi)^2 +\lambda^2 + \beta^2},
\end{equation}
and the heuristic static $k_3$,
\begin{equation}
k^{\text{static}}_3 = -\frac{\beta^2 D}{D^2+\lambda^2 + \beta^2},
\end{equation}
exemplifies the static approximation's remarkable accuracy in the
Boltzmann limit of the neutrino asymmetry evolution.  Indeed, this
definition of the matter and collision-affected mixing angle
arises naturally from the QKEs with
\begin{eqnarray}
\label{yoyo} \sin^2 2 \theta_{m,D} & \equiv & \left|{\cal
U}^{-1}_{x3} \right|^2 + \left|{\cal U}^{-1}_{y3} \right|^2 =
-\frac{k_3}{D}, \nonumber \\ \cos^2 2 \theta_{m,D} &\equiv&
\left|{\cal U}^{-1}_{z3} \right|^2 = 1 + \frac{k_3}{D},
\end{eqnarray}
that are quite visibly related to the projections of the
precession axis $\hat{q}_3$ in the fixed ${\cal S}$ basis, as
shown schematically in Fig.\ \ref{angle}. We refer the reader to
Appendix \ref{projection} for a more detailed discussion.

The damping term $\gamma$ in Eq.\ (\ref{gamma}) is equivalent to
$d$, the real part of the complex eigenvalues as defined in Eq.\
(\ref{k1k2}), and the corollary
\begin{equation}
\Gamma = 2 \gamma + \xi,
\end{equation}
illustrates lucidly the dual role played by collisions: (i) $2
\gamma$ rapidly damps the oscillations, and (ii) $\xi$  drives the
reaction $\nu_{\alpha} \rightleftharpoons \nu_s$ to equilibrium in
the long run. The second order differential equation, Eq.\
(\ref{2nd}), evokes oscillations at a damping-affected frequency
\begin{eqnarray}
\label{affectednu} \nu^2 & = & \nu_0^2 - \gamma^2  \nonumber \\
&=& \lambda^2 + \beta^2 + k_3 D + \frac{3}{4} k_3^2,
\end{eqnarray}
that is identified with $\omega$ of Eq.\ (\ref{k1k2}), the
imaginary component of complex eigenvalue.  Thus collisions seem
to modify the oscillation frequency in two ways by, (i)
contributing to the effective mass to produce a natural frequency
that is larger than the $D$-free frequency evident in Eq.\
(\ref{nuzero}), and (ii) reducing the natural frequency through
the damping term $\gamma$, {\it \`{a} la} classical oscillators by
Eq.\ (\ref{affectednu}). The end result, however, is that the two
effects seem to negate each other to some extent.  This is as yet
not very well understood.

\section{Non-adiabaticity}
\label{nonadiabaticity}

Significant variations in some or all of the parameters $D$,
$\lambda$ and $\beta$ over a characteristic time scale $\delta
t \sim (D^2 + \lambda^2 + \beta^2)^{-\frac{1}{2}}$ generally result
in non-unitary mapping between adjacent instantaneous diagonal
bases ${\cal S}_q(D,\,\lambda,\, \beta)$ and ${\cal S}_q(D+
\frac{\partial D}{\partial t} \delta t,\,\lambda+ \frac{\partial
\lambda}{\partial t} \delta t,\,\beta + \frac{\partial
\beta}{\partial t} \delta t)$, i.e.,
\begin{equation}
{\cal U}(D+ \frac{\partial D}{\partial t} \delta t,\, \lambda +
\frac{\partial \lambda}{\partial t} \delta t,\, \beta +
\frac{\partial \beta}{\partial t} \delta t) {\cal U}^{-1} (D,\,
\lambda,\, \beta) \simeq \text{diag}(1,1,1) - {\cal U}
\frac{\partial {\cal U}}{\partial t}^{-1} \delta t,
\end{equation}
where
\begin{equation}
{\cal U}(D+ \frac{\partial D}{\partial t} \delta t,\, \lambda +
\frac{\partial \lambda}{\partial t} \delta t,\, \beta +
\frac{\partial \beta}{\partial t} \delta t) \simeq {\cal U}(D,\,
\lambda,\, \beta) + \frac{\partial}{\partial t} {\cal U}(D,\,
\lambda,\, \beta) \delta t,
\end{equation}
and we have used $\frac{\partial}{\partial t}({\cal U}{\cal
U}^{-1}) = \frac{\partial {\cal U}}{\partial t}{\cal U}^{-1} +
{\cal U} \frac{\partial {\cal U}}{\partial t}^{-1}
=0$.\footnote{Predictably, the characteristic time $\delta t \sim
1/\nu_0 \simeq (D^2 + \lambda^2 + \beta^2)^{-\frac{1}{2}}$ is the
local {\it natural} oscillation length of the neutrino system as
defined in Sec.\ \ref{classical}, to be further discussed later in
Sec.\ \ref{rigorous}.} This corresponds to the non-adiabatic
regime where the computationally convenient decoupling of the
evolution of $Q_1$, $Q_2$ and $Q_3$ is invalid amidst sizeable
off-diagonal entries in the matrix ${\cal U} \frac{\partial {\cal
U}}{\partial t}^{-1}$ that essentially re-couple the named
variables in Eq.\ (\ref{qDE}). The effects of these off-diagonal
terms are most discernible near the resonance region where the
characteristic time $\delta t$ is a maximum. Full analytical
quantification of their contribution generally requires one to
inflate the homogeneous QKEs in Eq.\ (\ref{pqke}) into three
independent third order ordinary differential equations, which,
unfortunately, are not readily soluble even for a simple linear
$\lambda(t)$ profile, let alone three time-dependent parameters
$D$, $\lambda$ and $\beta$. However, the generic role played by
rapidly changing parameters may be understood through
consideration of the following simplified situation.

\subsection{A toy model}
\label{toymodel}

Suppose the neutrino ensemble evolves adiabatically up to time
$t_c$, at which $\lambda = \lambda_c$, $D = D_c$ and $\beta=
\beta_c$, such that
\begin{equation}
P_{\alpha}(t_0) \simeq {\cal U}^{-1}_{\alpha 3} (t_0, \lambda_c)
e^{ \int^{t_c}_{0} k_3(t')dt'} Q_3(0),
\end{equation}
with $\alpha = x,\, y,\, z$, and substantial damping is implicit.
Resonance crossing is instigated through the instantaneous
switching of $\lambda_c$ to $-\lambda_c$ at time $t_c$, after
which the system continues to undergo adiabatic evolution to time
$t$, that is,
\begin{eqnarray}
P_{\alpha}(t) &\simeq& \sum_{i,\, \beta} {\cal U}^{-1}_{yi}(t)
e^{\int^t_{t_c} k_i (t') dt'}{\cal U}_{i \beta}(t_c,-\lambda_c)
P_{\beta}(t_c) \nonumber \\
 &\simeq& \sum_{i,\, \beta} {\cal
U}^{-1}_{yi}(t) e^{\int^t_{t_c} k_i (t') dt'} \left[ {\cal U}_{i
\beta}(t_c,-\lambda_c) {\cal U}^{-1}_{\beta 3} (t_c, \lambda_c)
\right] e^{ \int^{t_c}_{0} k_3(t')dt'}  Q_3(0).
\end{eqnarray}
In the $D,\, |\lambda| \gg |\beta|$ limit, the term
$\sum_{\beta}[\cdots]$ evaluates explicitly to
\begin{equation}
\sum_{\beta} {{\cal U}_{i \beta}(t_c,-\lambda_c) {\cal
U}^{-1}_{\beta 3} (t_c, \lambda_c)}\doteq
    \left( -\frac{\sqrt{2}\beta_c \lambda_c}{D_c^2 +
    \lambda_c^2},\quad -\frac{\sqrt{2}\beta_c \lambda_c}{D_c^2 +
    \lambda_c^2},\quad
    1 - \frac{\beta_c^2}{D_c^2 + \lambda_c^2} \right),
\end{equation}
where we have used
\begin{equation}
\left.{\cal U}^{-1}\right|_{D \gg |\beta|}=
    \left( \begin{array}{ccc}
   \frac{1}{\sqrt{2}} \frac{\lambda}{|\lambda|} &
     \frac{1}{\sqrt{2}} \frac{\lambda}{|\lambda|}
     & \frac{\beta
    \lambda}{D^2 + \lambda^2} \\
    - \frac{i}{\sqrt{2}} & \frac{i}{\sqrt{2}} &
    -\frac{\beta D}{D^2 + \lambda^2} \\
    \frac{i \beta}{\sqrt{2}} \frac{D + i |\lambda|}{D^2 +
    \lambda^2} &  -\frac{i \beta}{\sqrt{2}}
    \frac{D - i |\lambda|}{D^2 + \lambda^2} & 1 \end{array}
    \right)  + {\cal O}(\beta^2),
\end{equation}
and
\begin{equation}
\left.{\cal U} \right|_{D \gg |\beta|} =
    \left( \begin{array}{ccc}
        \frac{1}{\sqrt{2}} \frac{\lambda}{|\lambda|} &
        \frac{i}{\sqrt{2}} & \frac{i \beta}{\sqrt{2}} \frac{D + i
        |\lambda|}{D^2 + \lambda^2} \\
         \frac{1}{\sqrt{2}} \frac{\lambda}{|\lambda|}&
        - \frac{i}{\sqrt{2}} & -\frac{i \beta}{\sqrt{2}} \frac{D - i
        |\lambda|}{D^2 + \lambda^2} \\
        \frac{\beta \lambda}{D^2 + \lambda^2}
        & \frac{\beta D}{D^2 + \lambda^2} & 1 \end{array} \right)
        + {\cal O}(\beta^2),
\end{equation}
so that to the lowest order in $\beta$,
\begin{eqnarray}
P_y(t) &\simeq& -\frac{\beta D}{D^2 + \lambda^2} e^{ \int^{t}_{0}
k_3(t')dt'}  Q_3(0) \nonumber \\ && \quad \quad- \frac{2\beta_c
\lambda_c}{D_c^2 + \lambda_c^2} e^{-\int^t_{t_c} D(t') dt'+
\int^{t_c}_0 k_3(t') dt'} \sin \left[\int^t_{t_c}
|\lambda(t')|dt'\right] Q_3(0), \nonumber
\\ P_z(t) &\simeq& e^{ \int^{t}_{0} k_3(t')dt'}  Q_3(0) + {\cal
O}(\beta^2).
\end{eqnarray}
The final step of letting $\lambda_c$ approach the vicinity of the
resonance, $\lambda_c \to D_c$, where $|\Delta \lambda| \simeq D$
is to be interpreted as the effective resonance width (see
Appendix \ref{projection}), completes this toy model,
\begin{equation}
\label{toypy}
\left. P_y(t)\right|_{\lambda_c\to D_c} \simeq
\left\{
    -\frac{\beta D}{D^2 + \lambda^2}
    - \frac{\beta_c}{D_c} e^{-\int^t_{t_c} D(t') + k_3(t') dt'}
    \sin \left[\int^t_{t_c} |\lambda(t')|dt'\right]
 \right\}P_z(t).
\end{equation}
Comparing Eqs.\ (\ref{pypzapprox}) and (\ref{toypy}), we find that
the first term in the latter is simply the adiabatic result in the
$D \gg |\beta|$ limit.  The remaining term quantifies oscillations
induced by the sudden change in $\lambda$ near the resonance (the
connection between oscillations and non-adiabatic effects was
first discussed in Ref.\ \cite{longpaper}). These ${\cal
O}(\beta)$ oscillations, although eventually exponentially damped
over a time scale of $\sim 1/D$, are most visible immediately
after resonance crossing, $t \stackrel{>}{\sim} t_c$, as the two
terms in Eq.\ (\ref{toypy}) may then be comparable in size.

\subsection{A more rigorous treatment}
\label{rigorous}

The aforementioned post-resonance oscillations may be shown to
arise formally from a more exact approach that incorporates the
actual, finite time derivatives $\frac{\partial D}{\partial t}$,
$\frac{\partial \lambda}{\partial t}$ and $\frac{\partial
\beta}{\partial t}$.  Following Ref.\ \cite{longpaper}, we
institute the complex variable
\begin{equation}
\tilde{P} \equiv P_x + i P_y,
\end{equation}
together with  its complex conjugate $\tilde{P}^*$, which, from
Eq.\ (\ref{pqke}), advances in time according to the equation
\begin{equation}
\label{tildepDE}
\frac{\partial \tilde{P}}{\partial t} = (-D + i
\lambda) \tilde{P} - i \beta P_z,
\end{equation}
and similarly for $\tilde{P}^*$.

The formal solution to Eq.\
(\ref{tildepDE}) is given by
\begin{eqnarray}
\label{tildepsoln}
\tilde{P}(t) =\left[\tilde{P}^*(t)\right]^* &=
& e^{-\int^t_0 (D - i \lambda) dt'}\tilde{P}(0)- i \int^t_0
e^{-\int^t_{t'} (D - i \lambda) dt''} \beta P_z dt'  \nonumber \\
&\simeq& -i \int^t_0 e^{-\int^{t}_{t'} (D - i \lambda) dt''} \beta
P_z  dt',
\end{eqnarray}
where we have used the initial conditions $P_x(0) \simeq P_y(0)
\simeq 0$ to establish the last approximate equality.  Note that
since we are working in the $D \gg |\beta|$ limit, even if these
initial values are not exactly vanishing, sufficient exponential
damping will nonetheless obliterate the term proportional to
$\tilde{P}(0)$ in time as discussed earlier in Sec.\
\ref{boltzmannlimit}.

Applying integration by parts on Eq.\
(\ref{tildepsoln}), we obtain
\begin{eqnarray}
\tilde{P}(t) &\simeq & -i \int^t_0 (D-i \lambda) e^{-\int^{t}_{t'}
(D - i \lambda) dt''} \frac{\beta P_z}{D-i \lambda} dt' \nonumber
\\
&=& - \left.\frac{i \beta P_z}{D - i \lambda} \right|_{t'=t} +
e^{-\int^t_0 (D - i \lambda) dt'} \left.\frac{i \beta P_z}{D - i
\lambda} \right|_{t'=0} + i \int^t_0 e^{-\int^t_{t'} (D - i
\lambda) dt''} \frac{d}{d t'} \left(\frac{\beta P_z}{D - i
\lambda} \right) dt'.
\end{eqnarray}
Substantial exponential damping again wipes out the second term,
such that $\tilde{P}(t)$ and $\tilde{P}^*(t)$ amalgamate to give
\begin{eqnarray}
\label{nonadpy}
P_y(t) &=& \frac{1}{2i} \left[\tilde{P}(t) -
\tilde{P}^*(t) \right] \nonumber
\\ &\simeq& - \frac{\beta D}{D^2 + \lambda^2} P_z(t)
 + \int^t_0 \frac{\beta}{\sqrt{D^2 + \lambda^2}} \left[\text{Re}
(\Xi) e^{-\int^t_{t'} D dt''}\sin \left( \int^t_{t'} \lambda
dt''\right) \right. \nonumber \\ &&  \hspace{7cm}  +
\left.\text{Im} (\Xi) e^{-\int^t_{t'} D dt''}\cos
\left(\int^t_{t'} \lambda dt''\right) \right] dt',
\end{eqnarray}
where $\text{Re}(\Xi)$ and $\text{Im}(\Xi)$ denote respectively
the real and imaginary parts of
\begin{equation}
\label{Xi} \Xi =  \frac{iD - \lambda}{\sqrt{D^2 + \lambda^2}}
\left\{ \left[ \frac{1}{\beta} \frac{\partial \beta}{\partial t'}
- \frac{1}{D - i \lambda} \left(\frac{\partial D}{\partial t'}  -
i\frac{\partial \lambda}{\partial t'}\right)  \right] P_z +
\frac{\partial P_z}{\partial t'} \right\}.
\end{equation}
Evidently, the first term in Eq.\ (\ref{nonadpy}) is simply the
adiabatic Boltzmann result in the $D \gg |\beta|$ limit.  The
accompanying integral is a close analogue of the ${\cal O}(\beta)$
oscillatory term found in the toy model in Eq.\ (\ref{toypy}), and
may be interpreted as a continuous sum of perpetuating
oscillations induced  by sizeable finite time
 derivatives $\frac{\partial D}{\partial
t}$, $\frac{\partial \lambda}{\partial t}$ and $\frac{\partial
\beta}{\partial t}$ at time $t'$ over the history of the evolution
(recall that the change in $\lambda$ is instantaneous in the toy
model).   Observe also that in the $D \gg |\beta|$ limit, the
quantity $\frac{\partial P_z}{\partial t} = \beta P_y$ is a
negligible ${\cal O}(\beta^2)$ term that may be readily verified
by iterating Eq.\ (\ref{nonadpy}).

Thus what remains in Eq.\
(\ref{Xi}) to first order in $\beta$ is in fact equivalent to
 elements in the matrix  ${\cal U}\frac{\partial {\cal U}}{\partial
t}^{-1}$ in Eq.\ (\ref{qDE}), that is \cite{bvw},
\begin{equation}
{\cal U} \frac{\partial {\cal U}}{\partial t}^{-1} =
    \left( \begin{array}{ccc}
            0 & 0 & - Z \\
            0 & 0 & - Z^* \\
            Z & Z^* & 0 \end{array} \right) + {\cal O}(\beta^2),
\end{equation}
where
\begin{equation}
Z = \frac{1}{\sqrt{2}}\frac{\beta}{\sqrt{D^2 + \lambda^2}} \times
\frac{iD - \lambda}{\sqrt{D^2 + \lambda^2}} \left[ \frac{1}{\beta}
\frac{\partial \beta}{\partial t} - \frac{1}{D - i \lambda}
\left(\frac{\partial D}{\partial t} - i\frac{\partial
\lambda}{\partial t}\right) \right],
\end{equation}
that were previously discarded in the adiabatic limit.

The integrand in Eq.\ (\ref{nonadpy}) is largest at $\lambda \sim
0$, i.e., in the proximity of a resonance.  In the context of
lepton number generation, we expect the term $\frac{\partial
\lambda}{\partial t}$ to dominate over $\frac{\partial D}{\partial
t}$ and $\frac{\partial \beta}{\partial t}$ in the
region of exponential growth (i.e., $\frac{d L_{\alpha}}{dt}
\propto L_{\alpha}$), where, by Eq.\ (\ref{ap}), the rate of
change of $\lambda$ is also correspondingly rapid.  Hence, Eq.\
(\ref{Xi}) reduces to
\begin{equation}
|\Xi| \sim \frac{1}{D} \left|\frac{\partial \lambda}{\partial t}
\right|_{\lambda \sim 0} \simeq \frac{1}{|\Delta \lambda|}
\left|\frac{\partial \lambda}{\partial t} \right|_{\lambda \sim
0},
\end{equation}
from which we identify $1/|\Xi|$ as the {\it physical} resonance
width, given $|\Delta \lambda| \simeq D$ is the collision-affected
resonance width in phase space.  The corresponding characteristic
time of the phase factor $\exp [-\int^t_{t'} (D - i \lambda)
dt'']$ is
\begin{equation}
\delta t \sim \frac{1}{\sqrt{D^2 + \lambda^2}} \sim \frac{1}{D},
\end{equation}
to be interpreted in the $D \gg |\beta|$ limit as the local
natural oscillation length of the neutrino system.  Thus the
degree of non-adiabaticity may be quantified by a {\it
collision-affected adiabaticity parameter}
\begin{equation}
\Upsilon \equiv |\Xi| \delta t \sim \frac{1}{D^2}
\left|\frac{\partial \lambda}{\partial t} \right|_{\lambda \sim
0},
\end{equation}
defined as a comparison between the physical resonance width and
the natural oscillation length of the neutrino system at $\lambda
\sim 0$, such that the condition $\Upsilon \ll 1$ denotes an ideal
adiabatic process, and vice versa.  The role of $\Upsilon$ is
illustrated in Fig.\ \ref{nonad} which shows $P_y(t)$ as a
function of $\lambda(t)$, where $\lambda(t)$ varies linearly with
time from $+ \lambda$ at $t = 0$ [initial conditions: $P_x(0)
\simeq P_y(0) \simeq 0$, and $P_z(0) \simeq 1$ ] to $- \lambda$,
for six different matter density gradients $\frac{\partial
\lambda}{\partial t}$. The parameters $D$ and $\beta$ are kept
constant for simplicity, and $P_z(t) \simeq 1$ throughout the
evolution.

Observe in Fig.\ \ref{nonad} that a sufficiently  adiabatic
process always returns a negative $P_y$. Substantial non-adiabatic
oscillations, however, can carry $P_y$ periodically across the
zero mark immediately after resonance crossing. From the
perspective of neutrino asymmetry evolution, occurrence around the
bulk of the momentum distribution could conceivably alternate the
sign of the rate $\frac{d L_{\alpha}}{dt}$ in Eq.\
(\ref{approxdldt}) in a cyclic manner, leading possibly to
oscillations in the integrated variable $L_{\alpha}$ in the region
of exponential growth.  This effect has been studied numerically in
Ref.\ \cite{p&r} (see also Ref.\ \cite{otherLoscs}).
Although a regime of rapid oscillatory
lepton number evolution has yet to be completely confirmed \cite{p&r},
non-adiabatic effects do indicate that their existence is likely.
This reinforces conclusions first reached in Ref.\ \cite{longpaper}.

\section{Conclusion}
\label{conclusion}

The Boltzmann limit of the relic neutrino asymmetry evolution
represents a phase in which the rate of the said evolution is
dependent only on the instantaneous neutrino and antineutrino
distribution functions. An associated evolution equation may be
extracted from the exact QKEs in the adiabatic limit where the
matter potential and collision rate both vary slowly with time, in
alliance with adequate collision-induced damping which serves to
wipe out the coherence history of the system.  In the course of
the derivation, we have ascribed precise physical and/or
geometrical meanings for the adiabatic approximation and the many
quantities arising therein that were previously lacking.  These
original interpretations allow one to easily visualise the sharply
contrasting behaviours exhibited by the neutrino ensemble across
an MSW resonance in the collision-dominated and the coherent
oscillatory regimes respectively.

The time development of the individual neutrino and antineutrino
ensembles, in particular the difference in $\nu_{\alpha}$ and
$\nu_s$ distribution functions, is mimicked by a damped classical
system oscillating about a midpoint which contemporaneously
``decays'' with time. We identify this decay constant with the
``rate constant'' that couples to the evolution equation for
$(N_{\alpha} - N_s)$.  Thus the Boltzmann limit is revealed to
consist of approximating the intrinsically oscillatory quantity
$(N_{\alpha} - N_s)$ to some stable average $(N_{\alpha} - N_s)_0$
which evolves in time in the same manner as $(N_A - N_B)$ in the
generic reaction $A \rightleftharpoons B$ with identical forward
and reverse rate constants.  The approximation's validity is
guaranteed for small-amplitude oscillations through substantial
damping and/or matter suppression. The pertinent reaction rate
constant is essentially a product of the decoherence rate and the
neutrino ensemble's matter and collision-affected mixing angle,
the latter of which is a new quantity identified here for the
first time.  Collisions therefore play two roles: (i) to quickly
damp the oscillations, and (ii) to drive the ``static'' reaction
$\nu_{\alpha} \rightleftharpoons\nu_s$ to equilibrium over an
extended period of time.

Significant time variations in the damping and mixing parameters,
i.e., non-adiabatic evolution, are shown to induce comparatively
large amplitude oscillations in the system immediately after
resonance crossing. The degree of adiabaticity is quantified by an
 eponymous parameter first introduced in this work, whose
 role in collision-affected neutrino oscillation dynamics
 parallels
 that of its more common MSW-style counterpart in completely coherent scenarios.
   Substantial oscillations in the bulk of the momentum distribution
   may lead to periodic
alternation in the sign of the quantity $\frac{d L_{\alpha}}{d
t}$, and are thus a prime suspect for the generation of rapid
oscillations observed by others in the course of the asymmetry
evolution.

 Ultimately we would like to improve on
the adiabatic Boltzmann approximation, and results from the
present work have put us on better grounds for its accomplishment.
The classical oscillator analogy, for instance, seems to suggest a
procedure by which to correct for phase dependence in the
asymmetry evolution.  This and other avenues await to be explored
in the future.

\acknowledgments{This work was supported in part by the Australian
Research Council and in part by the Commonwealth of Australia's
postgraduate award scheme.  YYYW would like to thank S. N. Tovey
for his very inspiring Part I Classical Mechanics lectures.}

\appendix
\section{Cubic polynomial}
\label{cubic}

We present further information on the derivation of Eqs.\
(\ref{discriminant}) to (\ref{omega}), and other conditions
arising from the characteristic equation in Eq.\
(\ref{characteristic}) scattered throughout Sec.\
{\ref{adiabatic}.

 The cubic polynomial under investigation is the characteristic equation
 of the $3 \times 3$ matrix ${\cal K}$ of Eq.\ (\ref{pqke}),
\begin{equation}
\label{poly}
x^3 + 2D x^2 + (D^2 + \lambda^2 + \beta^2) x +
\beta^2 D = 0,
\end{equation}
where the coefficients $a_1 = 2D$, $a_2=D^2+\lambda^2 + \beta^2$,
and $a_3 = \beta^2 D$ are real and positive.  We seek three
solutions $k_1$, $k_2$ and $k_3$, some of which may be complex, to
Eq.\ (\ref{poly}), that is,
\begin{eqnarray}
& (x - k_1)(x - k_2)(x - k_3) =  0 \nonumber \\ \Rightarrow & x^3
- (k_1 + k_2 + k_3) x^2 + (k_1 k_2 + k_2 k_3 + k_3 k_1) x - k_1
k_2 k_3 = 0,
\end{eqnarray}
with
\begin{eqnarray}
\label{coeff} k_1 + k_2 + k_3 & = & -2D, \nonumber \\ k_1 k_2 +
k_2 k_3 + k_3 k_1 & = & D^2 + \lambda^2 + \beta^2, \nonumber
\\ k_1 k_2 k_3 & = & - \beta^2 D.
\end{eqnarray}
Equation (\ref{coeff}) shows plainly that any complex roots must
occur as a conjugate pair, and the remaining one is necessarily
real. For this we introduce the following parameterisation,
\begin{equation}
\label{pilot}
k_{1,\,2} =  -d \pm i \omega,
\end{equation}
where the quantities $d$ and $\omega$ are defined to be real and
positive.  Note that the definition of $\omega$ may be extended to
include imaginary numbers, in which case $k_1$ and $k_2$ are
simply two distinct, real roots.  Simple algebraic manipulation of
Eqs.\ (\ref{coeff}) and (\ref{pilot}) leads to
\begin{eqnarray}
k_3 &=& -\frac{\beta^2 D}{d^2 + \omega^2}, \label{pencil} \\ d & =
& D+\frac{k_3}{2}, \label{paper}
\\ \omega^2 &=& \lambda^2 +
\beta^2 + k_3 D + \frac{3}{4} k_3^2, \label{rubber}
\end{eqnarray}
which are identically Eqs.\ (\ref{k3}) to (\ref{omega}) in the
main text.  The quantity
\begin{equation}
\label{ubiquitous}
\Delta = - \left( k_1 - k_2 \right)^2 \left(k_2
- k_3 \right)^2 \left(k_3 - k_1 \right)^2,
\end{equation}
known as the discriminant, thus characterises the nature
of the three roots:
\begin{enumerate}
\item $\Delta > 0$.  One root is real and the other two a
complex conjugate pair.
\item $\Delta = 0$. All roots are real and at least two are equal.
\item $\Delta < 0$.  All roots are real and distinct.
\end{enumerate}
These shall be labelled conditions 1, 2 and 3 hereafter. The
general form of Eq.\ (\ref{ubiquitous}) in terms of the
coefficients $a_1$, $a_2$ and $a_3$ of the cubic polynomial may be
found in any standard mathematical handbook, that is,
\begin{equation}
\Delta  =  4 Q^3 + 27 R^2,
\end{equation}
with
\begin{eqnarray}
Q &=& \frac{3 a_2 - a_1^2}{3},\nonumber \\ R &=& \frac{9 a_1 a_2 -
27 a_3 - 2 a_1^3}{27}.
\end{eqnarray}
  Thus the discriminant of Eq.\
(\ref{poly}) is equivalently
\begin{equation}
\label{empire}
\Delta = 4 \beta^6 - \beta^4 D^2 + 12 \beta^4
\lambda^2 - 20 \beta^2 D^2 \lambda^2 + 4 D^4 \lambda^2 + 12
\beta^2 \lambda^4 + 8 D^2 \lambda^4 + 4 \lambda^6.
\end{equation}
The reader can easily verify that condition 1 eventuates if the
requirement $|\lambda| \gg D$ is met for all $\beta$, and
similarly for the case $|\beta| \gg D$ with a variable $\lambda$.
The $D \gg |\lambda|,\, |\beta|$ affair is a trifle more
intricate.  We begin by keeping only
terms up to order $(\beta/D)^4$ and $(\lambda/D)^4$ in Eq.\
(\ref{empire}).  Then condition 1 is attained if the inequality
\begin{equation}
\label{casualty} 8 \left( \frac{\lambda}{D} \right)^4 + \left[4 -
20 \left(\frac{\beta}{D} \right)^2\right] \left( \frac{\lambda}{D}
\right)^2 - \left(\frac{\beta}{D} \right)^4 > 0,
\end{equation}
holds.  Given that the above quadratic has solutions at
\begin{eqnarray}
4 \left(\frac{\lambda}{D}\right)^2 &=& -1 + 5 \left(
\frac{\beta}{D}\right)^2 \pm \sqrt{1 - 10 \left( \frac{\beta}{D}
\right)^2 + 27 \left(\frac{\beta}{D} \right)^4} \nonumber \\
&\simeq& -1 + 5 \left( \frac{\beta}{D}\right)^2 \pm \left[1 - 5
\left( \frac{\beta}{D} \right)^2 + \left(\frac{\beta}{D} \right)^4
+ {\cal O}\left(\frac{\beta^6}{D^6} \right)\right] \nonumber
\\ &\simeq& \left(\frac{\beta}{D}\right)^4, \quad \text{and} \quad
- 2 + {\cal O}\left( \frac{\beta^2}{D^2}
                           \right),
\end{eqnarray}
the inequality in Eq.\ (\ref{casualty}) may be similarly expressed
in terms of two conditions
\begin{eqnarray}
4 \left(\frac{\lambda}{D}\right)^2 &>&
\left(\frac{\beta}{D}\right)^4, \nonumber \\ 4
\left(\frac{\lambda}{D}\right)^2 &<& - 2 + {\cal O}\left(
\frac{\beta^2}{D^2}
                           \right).
\end{eqnarray}
Naturally, the second statement is never true. Hence, in the limit
$D \gg |\lambda|,\, |\beta|$, the discriminant is positive only
for
\begin{equation}
\lambda^2 > \frac{\beta^4}{4 D^2},
\end{equation}
as reported in Eq.\ (\ref{recipe}) in Sec.\ \ref{zerolambda}.

\section{Matter and collision-affected mixing angle}
\label{projection}

We begin our discussion by putting forward a conjecture, that the
mixing angle $\cos 2 \theta_{m,D}$, collision-affected or
otherwise, is the normalised $z$-component of the precession axis
$\hat{q}_3$ (see Fig.\ \ref{basis} for orientation), i.e.,
\begin{equation}
\label{cos} \cos^2 2 \theta_{m,D} \equiv \left|{\cal U}^{-1}_{z3}
\right|^2,
\end{equation}
where ${\cal U}^{-1}$ is the transformation matrix given by Eqs.\
(\ref{u-1matrix}) and (\ref{normalisation}).  It follows that the
remainder
\begin{equation}
\label{sin} \sin^2 2 \theta_{m,D}  \equiv  \left|{\cal
U}^{-1}_{x3} \right|^2 + \left|{\cal U}^{-1}_{y3} \right|^2,
\end{equation}
is generically projected onto the $xy$-plane.  In the limit $D=0$,
the unit vector $\hat{q}_3$ is parallel to the matter potential
vector ${\bf V}$ with no projection on the $y$-axis. Equations
(\ref{cos}) and (\ref{sin}) thereby reduce to the household
expressions readily obtainable from Eq.\ (\ref{potential}).

The general collision-affected case is somewhat more convoluted.
Installing the matrix elements from Eqs.\ (\ref{u-1matrix}) and
(\ref{normalisation}), Eq.\ (\ref{sin}) evaluates explicitly to
\begin{equation}
\sin^2 2 \theta_{m,D} = \frac{k_3^2 \left[\lambda^2 +
(D+k_3)^2\right]}{\lambda^2 k_3^2 + (\beta^2 + k_3^2) (D+k_3)^2}.
\end{equation}
Equations (\ref{k3}) to (\ref{omega}) combine to produce
\begin{equation}
k_3 = - \frac{ \beta^2 D }{ (D + k_3)^2 + \lambda^2 + \beta^2 },
\end{equation}
which permits us to
make the substitution $k_3 (D+k_3)^2 = - [\beta^2 D +
k_3(\lambda^2 + \beta^2)]$ to arrive at
\begin{equation}
\label{dodo} \sin^2 2 \theta_{m,D} = - \frac{k_3}{D} =
\frac{\beta^2}{(D+k_3)^2 + \lambda^2 + \beta^2}.
\end{equation}
We identify the denominator as, in the classical harmonic
oscillator vernacular, the ``natural frequency'' $\nu_0^2$ given
by Eq.\ (\ref{nuzero}) (with $k_3 = -\xi$).  Thus Eq.\
(\ref{dodo}), with its lofty QKE origin, is entirely equivalent to
the heuristically derived matter and collision-affected mixing
angle in Eq.\ (\ref{Dmixing}).

In the $D \gg |\beta|$ limit, Eq.\ (\ref{dodo}) peaks at $\lambda
=0$ with the value $\beta^2/D^2$, with resonance width $|\Delta
\lambda| \simeq D$. This is to be compared with the familiar $D=0$
case in which the nominal resonance width $|\Delta \lambda| \simeq
|\beta|$ corresponds to $\sin^2 2\theta_m$ attaining a maximum
value of unity at resonance.

\begin{figure}[htp]
\caption{Schematic representation of the polarisation ${\bf P}$,
the matter potential vector ${\bf V}$, and the instantaneous
diagonal basis ${\cal S}_q$ in relation to the fixed coordinate
system ${\cal S}$.  The authors warn that some features have been
exaggerated for pictorial clarity.
Note also that the unit vectors
$\hat{q}_1$ and $\hat{q}_2$ are complex vectors. \label{basis}}

\caption{Schematic representation of the precession axis
$\hat{q}_3$ for the case $|\beta| \stackrel{>}{\sim} D$ as a
function of the matter potential vector ${\bf V}$.  The tags
$(\beta,\,0,\,\lambda)$ show the pertinent values of ${\bf V}$.
\label{smalldq3}}

\caption{Schematic representation of the precession axis
$\hat{q}_3$ for the case $D \gg |\beta|$ as a function of the
matter potential vector ${\bf V}$.  The tags
$(\beta,\,0,\,\lambda)$ show the pertinent values
of ${\bf V}$.
\label{bigdq3}}

\caption{Evolution of the effective total lepton number
$L^{(\alpha)}$, where $\alpha = \mu,\, \tau$, for $\nu_{\alpha}
\leftrightarrow \nu_s$ oscillation parameters $\Delta m^2 = -
0.01\, \text{eV}^2$ and $\sin^2 2 \theta_0 = 10^{-7}$.  The dashed
and dotted lines represent results from numerically integrating
Eqs.\ (\ref{approxdldt}) and (\ref{sterile}) using the exact
eigenvalue $k_3$ and the heuristically derived
$k_3^{\text{static}}$ respectively.   These are juxtaposed with
the solution to the exact QKEs [Eq.\ (\ref{qkes})] for the same
oscillation parameters (solid line).  \label{asym2}}

\caption{Evolution of the effective total lepton number
$L^{(\alpha)}$, where $\alpha = \mu,\, \tau$, for $\nu_{\alpha}
\leftrightarrow \nu_s$ oscillation parameters $\Delta m^2 = - 10\,
\text{eV}^2$ and $\sin^2 2 \theta_0 = 10^{-7}$.  The dashed and
dotted lines represent results from numerically integrating Eqs.\
(\ref{approxdldt}) and (\ref{sterile}) using the exact eigenvalue
$k_3$ and the heuristically derived $k_3^{\text{static}}$
respectively.   These are juxtaposed with the solution to the
exact QKEs [Eq.\ (\ref{qkes})] for the same oscillation parameters
(solid line). \label{asym3}}

\caption{Damped harmonic oscillator with a ``decaying''
oscillation midpoint --- schematic representation of the solution
to Eq.\ (\ref{3rd}). \label{osc}}

\caption{Schematic representation of the matter and
collision-affected mixing angle $2 \theta_{m,D}$.  The ordinary
matter-affected mixing angle is labelled $2 \theta_m$.
\label{angle}}

\caption{The variable $P_y(t)$ as a function of $\lambda(t)$,
where $\lambda(t)$ varies linearly with time from $+ \lambda$ to
$-\lambda$, for adiabaticity parameters $\Upsilon =0.25$ (top
left), $0.625$ (top right), $1.25$ (centre left), $3.125$ (centre
right), $6.25$ (bottom left), and $18.75$ (bottom right). The
parameters $D$ and $\beta$ are held constant at $D=0.02$ and
$\beta=0.001$, and $P_z(t) \simeq 1$ throughout the evolution. The
initial conditions are $P_x(0) \simeq P_y(0) \simeq 0$, and
$P_z(0) \simeq 1$. \label{nonad}}

\newpage
\begin{center}

\epsfig{file=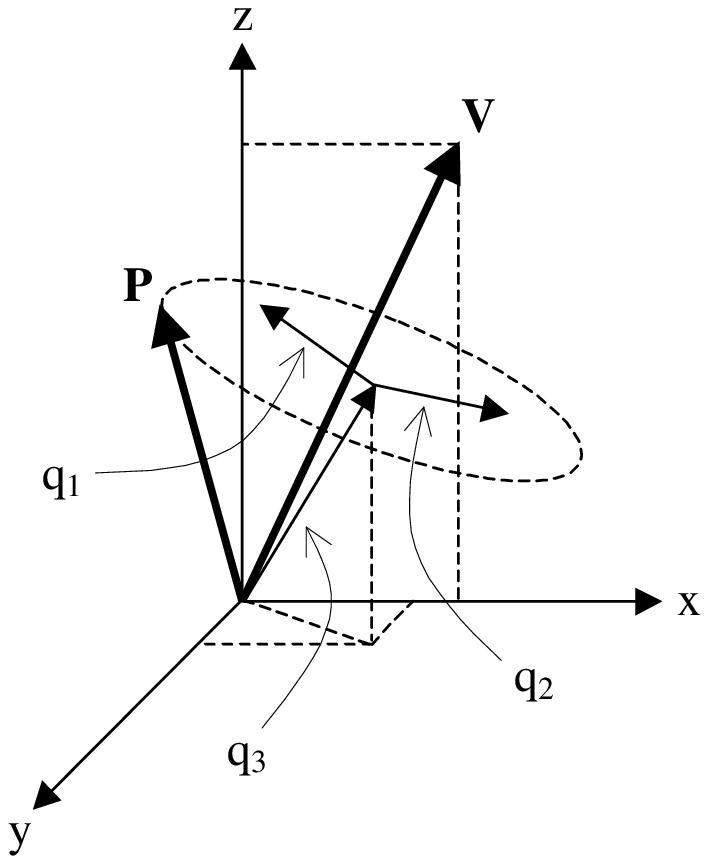,height=150mm}

\vspace{50mm} FIGURE \ref{basis}

\newpage
\epsfig{file=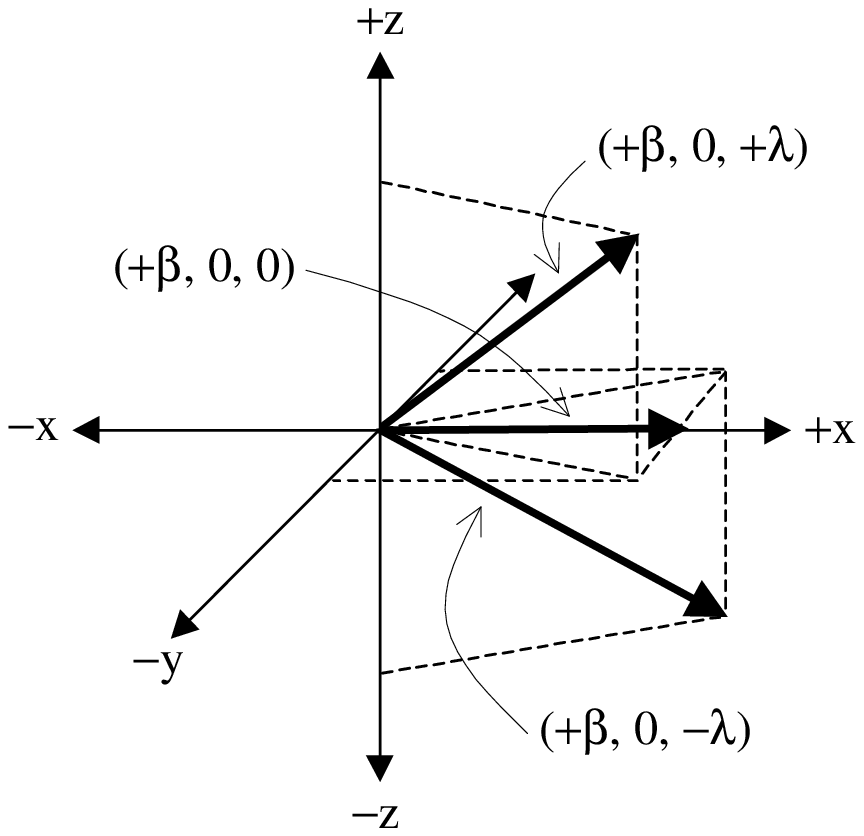,width=150mm}

\vspace{50mm} FIGURE \ref{smalldq3}

\newpage

\epsfig{file=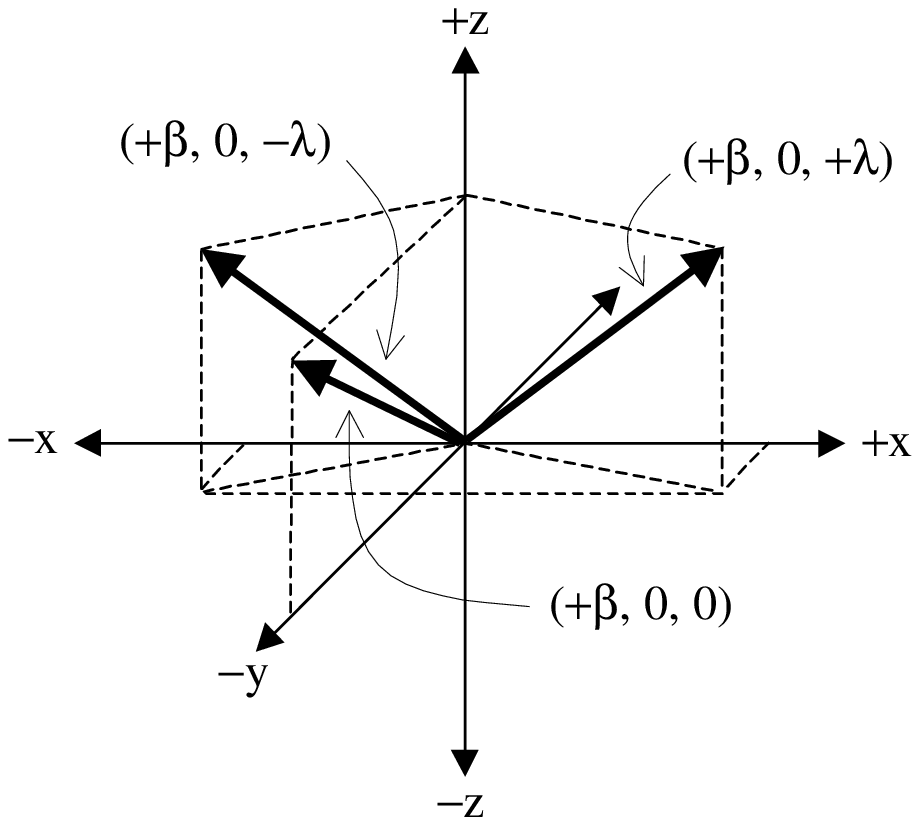,width=150mm}

\vspace{50mm} FIGURE \ref{bigdq3}

\newpage
\epsfig{file=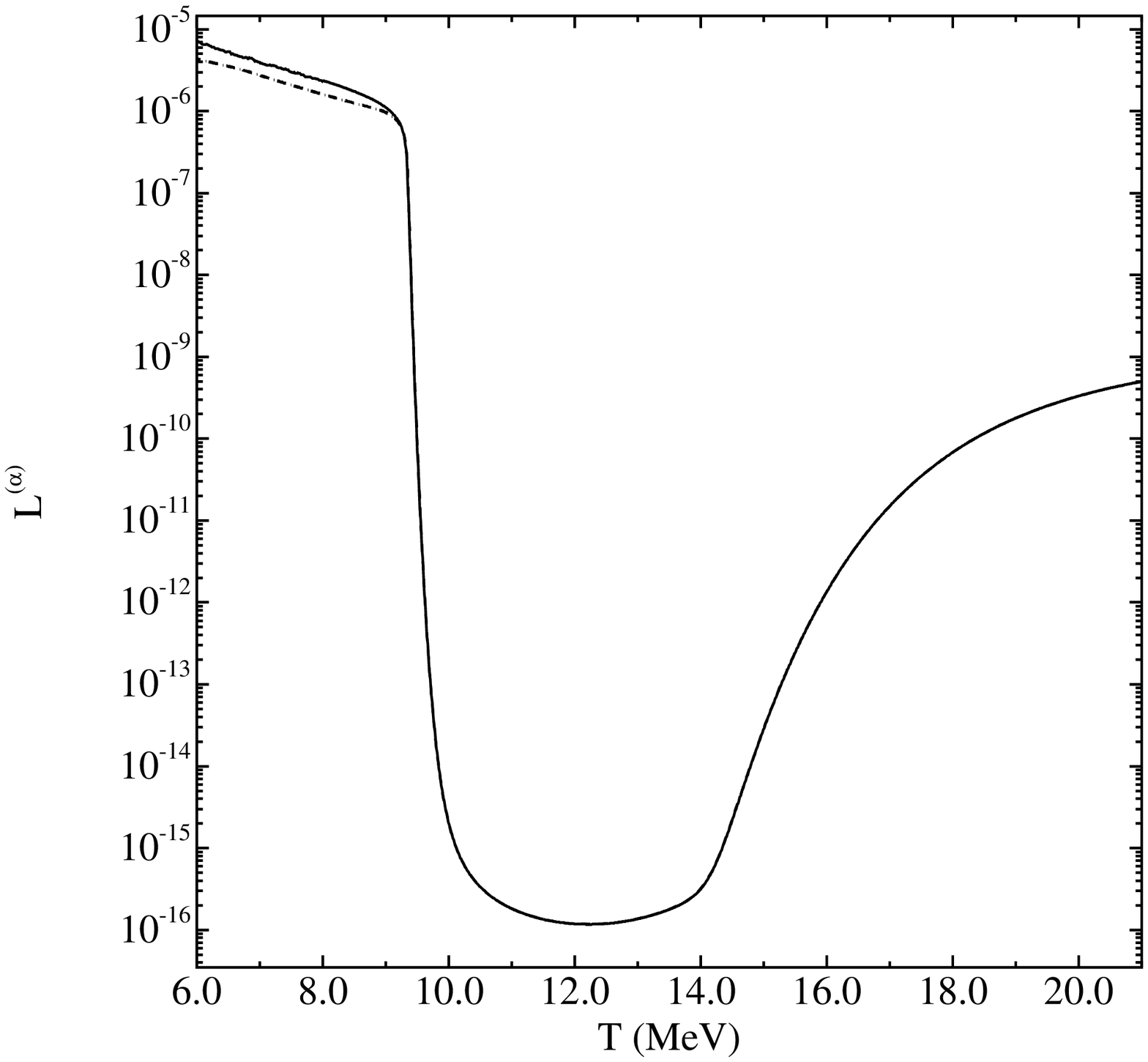,height=150mm,width=150mm}

\vspace{50mm} FIGURE \ref{asym2}

\newpage
\epsfig{file=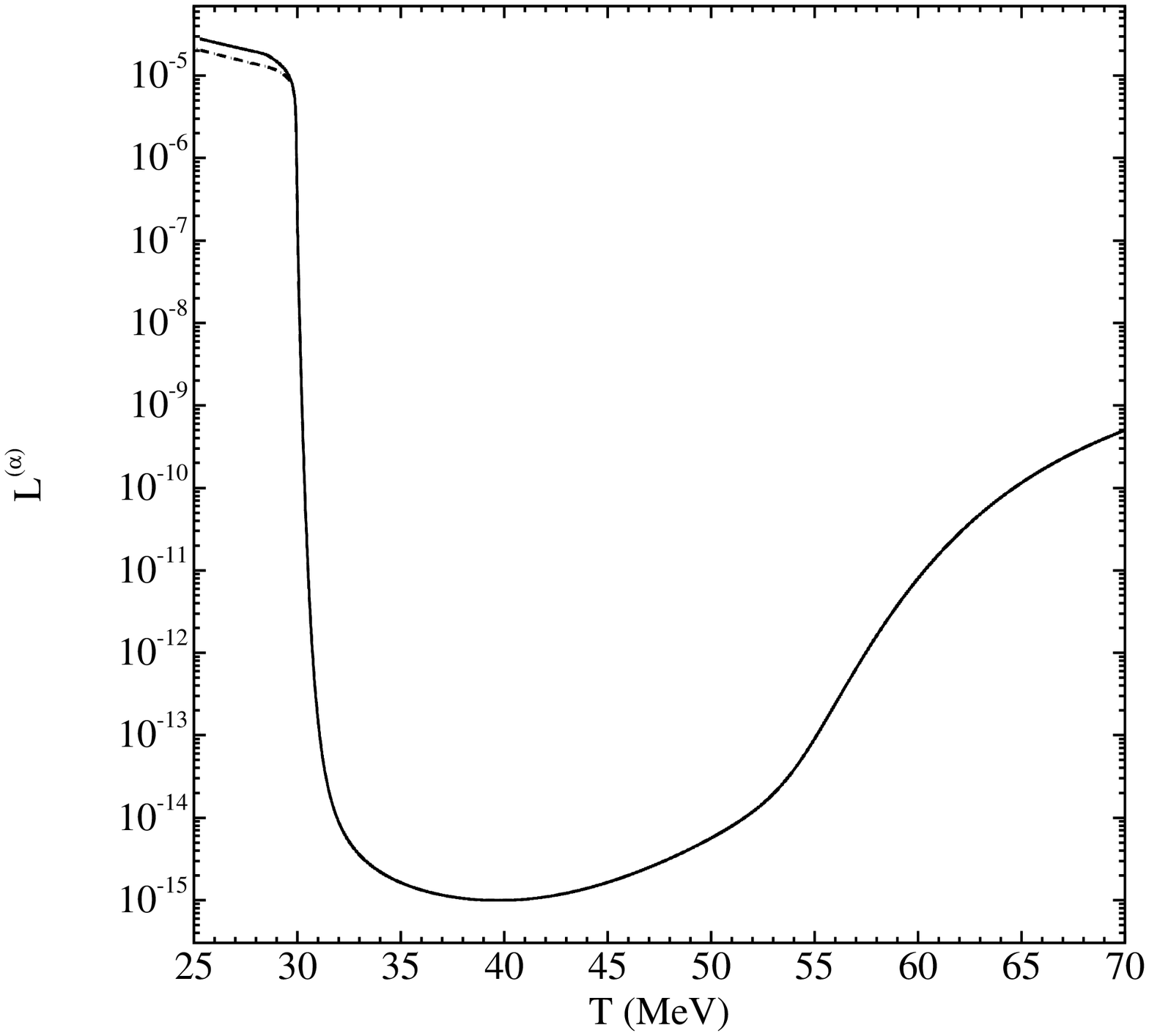,height=150mm,width=150mm}

\vspace{50mm} FIGURE \ref{asym3}

\newpage
\epsfig{file=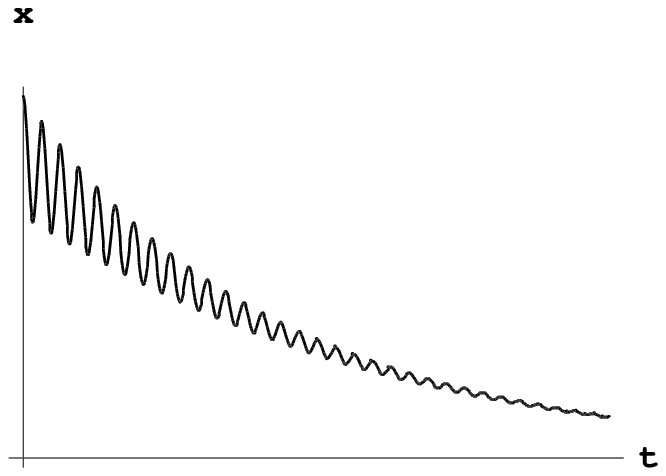,width=150mm}

\vspace{50mm} FIGURE \ref{osc}

\newpage
\epsfig{file=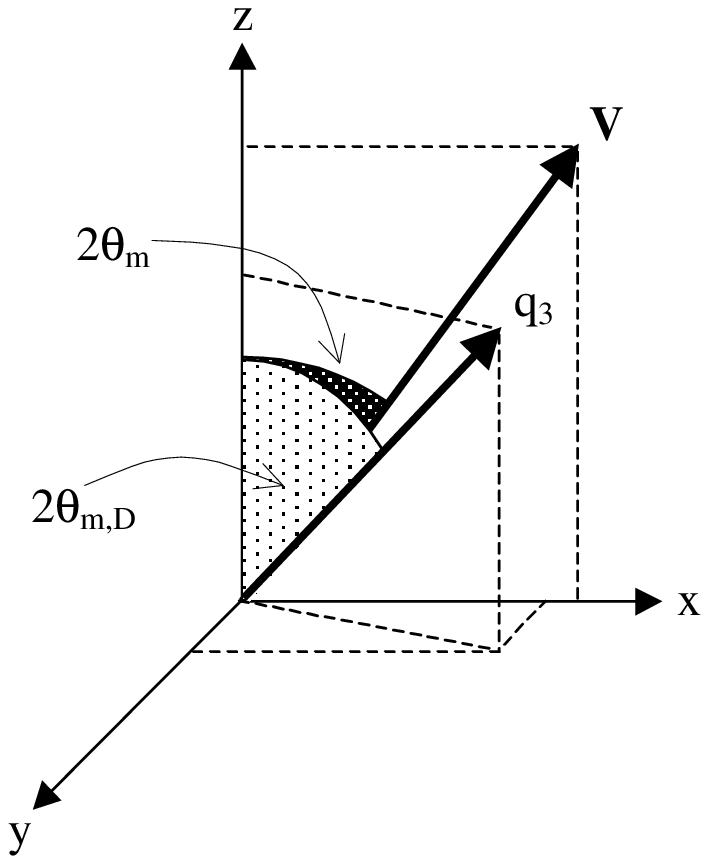,height=150mm}

\vspace{50mm} FIGURE \ref{angle}

\newpage
\epsfig{file=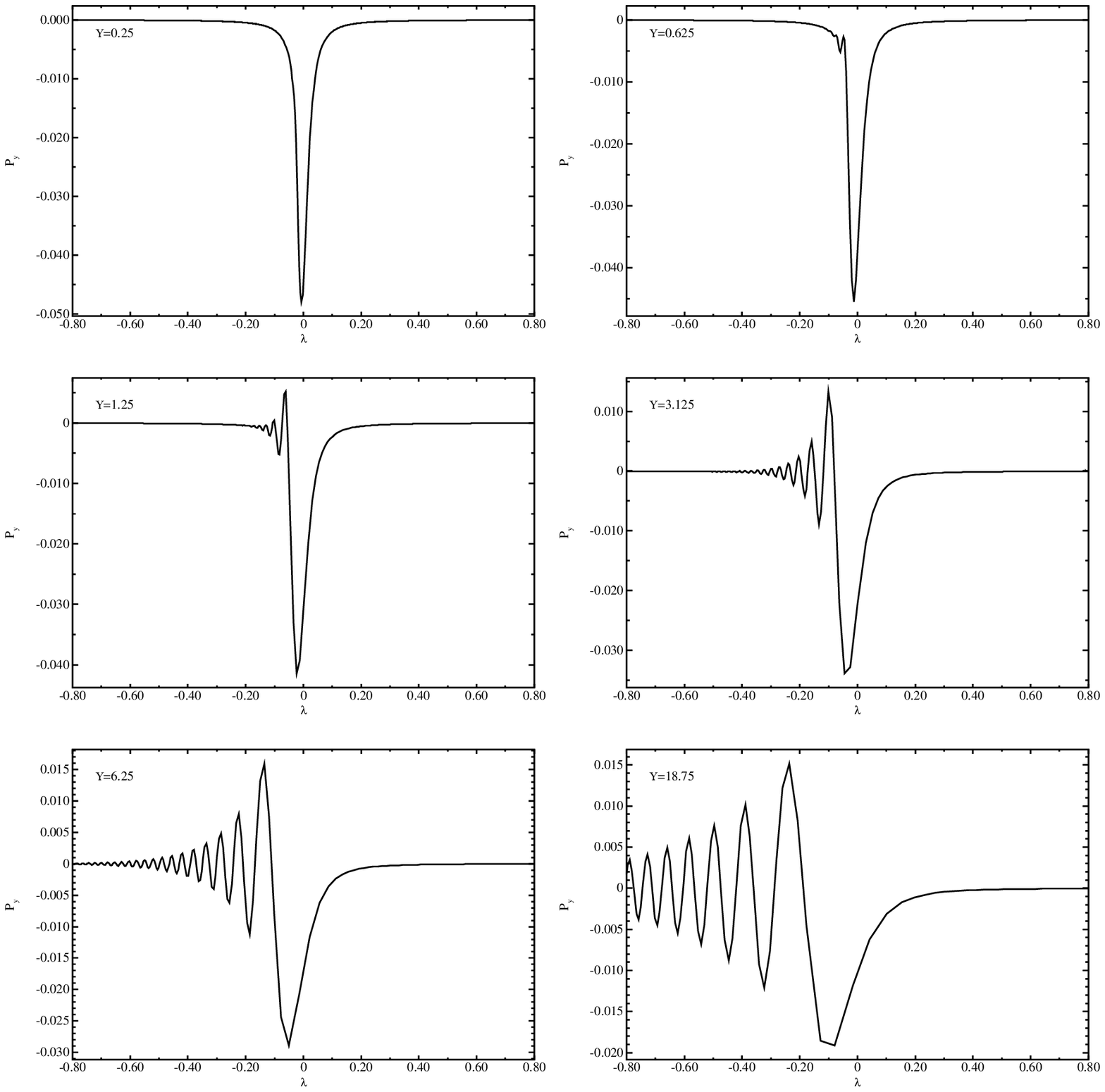,width=150mm}

\vspace{50mm} FIGURE \ref{nonad}
\end{center}
\end{figure}

\end{document}